\documentclass[a4paper,aps,prd,twocolumn,showpacs,eqsecnum,amsmath,amssymb,nofootinbib,superscriptaddress]{revtex4-1}
\usepackage{dcolumn}
\usepackage{bm}
\usepackage{graphics}
\usepackage{graphicx}
\usepackage{epsfig}
\usepackage{booktabs,multirow}
\usepackage{hhline}
\usepackage{slashed}
\usepackage{rccol}
\usepackage{mathrsfs}

\usepackage{xcolor}\colorlet{linkequation}{blue}
\usepackage[colorlinks=true, 
            linkcolor=red,   
            filecolor=red,   
            citecolor=green, 
            urlcolor=blue,
            hyperfootnotes=true]{hyperref}

\newcommand*{\SavedEqref}{}
\let\SavedEqref\eqref
\renewcommand*{\eqref}[1]{%
  \begingroup
    \hypersetup{
     linkcolor=linkequation,
      linkbordercolor=linkequation,
    }%
    \SavedEqref{#1}%
  \endgroup
}

\begin{document}
\newcommand{\newc}{\newcommand}
\renewcommand{\thefootnote}{\fnsymbol{footnote}}

\newc{\neutralino}{\widetilde\chi^0}
\newc{\chargino}{\widetilde\chi^{\pm}}
\newc{\squarkk}{\widetilde{q}_k}
\newc{\squarkl}{\widetilde{q}_l}
\newc{\tanb}{\tan\beta}
\newc{\gev}{\mbox{~GeV}}
\newc{\tev}{\mbox{~TeV}}

\title{Neutralino pair production via photon-photon collisions at the ILC}
\author{M.~Demirci}
\email{Corresponding author. mehmetdemirci@ktu.edu.tr}
\affiliation{Department of Physics, Karadeniz Technical University, 61080 Trabzon, Turkey}%
\author{A.~I.~Ahmadov}
\email{ahmadovazar@yahoo.com}
\affiliation{Department of Theoretical Physics, Baku State University, Z. Khalilov Street 23,\\ AZ-1148, Baku, Azerbaijan}%
\date{August 10, 2016}
\begin{abstract}
We provide complete one-loop predictions for the direct production of the neutralino pairs via photon-photon
collisions which appears for the first time at one-loop level in the minimal supersymmetric standard model at the International Linear Collider. We present a comprehensive investigation of the dependence of total cross section on the center-of-mass energy and the tan$\beta$ for three different scenarios based on gaugino/Higgsino fractions of the neutralinos. Furthermore, by investigating the behavior with the most relevant parameters $\mu$ and $M_2$, we exhibit regions of the parameter space where the enhancement of the cross section is large enough to be detectable at linear colliders. Our analysis shows that the corresponding production cross section reaches its highest values when the lightest neutralino $\widetilde{\chi}_{1}^0$ has a dominant Higgsino component and/or the next-to-lightest neutralino $\widetilde{\chi}_{2}^0$ has a dominant gaugino component.
\end{abstract}


\maketitle

\section{\bf Introduction}
Supersymmetry (SUSY) (see, e.g., Refs. \cite{Haber,Nilles,wss}) is a strongly motivated candidate for physics beyond the standard model (SM). It protects the Higgs vacuum expectation value without unnatural fine-tuning of the theory parameters, allows unification of gauge couplings at a high-energy scale, and offers  a candidate for dark matter (DM) postulated to explain astrophysical observations \cite{DM}. The minimal supersymmetric standard model (MSSM), one of the most widely studied and well-motivated SUSY frameworks, keeps the  number of new fields and couplings to a minimum. In  contrast  to  the  single  Higgs  doublet  of  the SM, the MSSM contains two Higgs doublets, which in the \textit{CP} conserving case leads to a physical spectrum consisting of the light/heavy \textit{CP}-even Higgs bosons $h^0$/$H^0$, a \textit{CP}-odd Higgs boson $A^0$, and a couple of charged Higgs bosons $H^\pm$. The experimentally observed Higgs with mass of about 125 GeV~\cite{Higgs_ATLAS,Higgs_CMS} can naturally be interpreted as the light or heavy \textit{CP}-even MSSM Higgs boson~\cite{Heinemeyer,Scopel,Djouadi}. The MSSM also predicts many new particles such as sleptons, squarks, four neutralinos $\neutralino_{i}$, and two charginos $\chargino_j$\footnote{The neutralinos/charginos are the mass eigenstates formed from the linear superposition of the neutral/charged superpartners of the electroweak gauge bosons
and Higgs doublets (the so-called gauginos and Higgsinos, respectively).}. In supersymmetric models with $R$-parity~\cite{Fayet} conservation, the lightest
neutralino $\neutralino_{1}$ is commonly assumed to be a weakly
interacting massive particle, which is consistent with the
observations of a DM candidate (for a review, see, e.g., Refs. \cite{Jungman, Griest}) in
the form of the lightest supersymmetric particle (LSP). It also
emerges as the final particle of the decay chain of each  the supersymmetric particles (or sparticles, for short)
and escapes from the detector without interacting. If the lightest neutralino $\neutralino_{1}$ is indeed the LSP, supersymmetric dark matter is suggested to be detected directly in present or forthcoming experiments~\cite{Arnowitt}. Accordingly, a detailed investigation of the lightest neutralino is essential for providing  information regarding the phenomenological and theoretical aspects of SUSY.

One of the most important goals of many colliders, particularly the LHC, is to discover sparticles (or any other sign of physics beyond the SM). Although SUSY retains strong theoretical arguments in its favor, no evidence of sparticles has yet been found from many searches for SUSY carried out at the LHC. These experimental searches which mainly focus on the production of the strongly interacting sparticles (such as squarks and gluinos) have pushed up the mass limits of these particles into the TeV region, depending on details of the assumed parameters~\cite{sqgl_ATLAS1,sqgl_ATLAS2,sqgl_CMS1,sqgl_CMS2}. Unlike the colored sparticles, the bounds on masses of sleptons, charginos, and neutralinos are relatively weak, particularly for the region of the compressed spectrum. On the other hand, naturalness suggests that neutralinos, charginos, and third-generation sparticles must be sufficiently light (a few hundreds of GeV) to produced at the TeV scale~\cite{Chan}. The bound on the mass of the neutralino $\neutralino_{1}$ is given by $m_{\neutralino_1} \gtrsim 46 \gev$ at 95\% CL, derived from the lower bound on the chargino mass in the MSSM with gaugino mass unification at the large electron positron collider (LEP) \cite{Abdallah}. In the constrained MSSM including both sfermion and gaugino mass unification, however, the bound reaches to well above 100 GeV from the powerful constraints set by the recent LHC data \cite{PDG}. A hunt related to the discovery (or exclusion) of any sparticles is still ongoing at the LHC.

In anticipation of much heavier colored sparticles, therefore, we are led to consider a more challenging search strategy, namely, the SUSY signals only from the electroweak sector, the neutralinos and  charginos. In scenarios where masses of colored sparticles are larger than a few TeV, the direct production of neutralinos and charginos can be the dominant process of SUSY at the LHC. However, this production
mode at the LHC suffers from both relatively small rates and large SM backgrounds~\cite{Han,HBaer,Demirci3}. This motivates the complementary experiments at future linear colliders with low backgrounds and easy signal reconstruction. Particularly, a linear collider would be best suited for producing the lighter sparticles. Linear-collider experiments could focus on one type of sparticle at a time, measuring their properties precisely enough to detect the symmetry of SUSY and to reveal the supersymmetric nature of DM.

Recently, there has been a proceeding effort for the International Linear Collider (ILC) designed to give facilities for electron-positron, electron-electron, and photon-electron collisions. The main goal at the ILC is to be a good complementarity of the LHC results and also open new windows in the phenomenological part of the physics beyond SM such as SUSY, little Higgs models, and extra gauge bosons. The ILC compared to the LHC has a cleaner background, and it is possible to extract the new physics signals from the background more easily. The ILC is the most advanced design as a future lepton collider laid out for the center-of-mass energy range of $\sqrt{s} =$ 90-1000 GeV \cite{ILC1,ILC2,eeLC}. The Compact Linear Collider (CLIC) is also another project which aims to work on a concept for a machine to collide electrons and positrons head on at energies up to several TeV. In case a drive beam accelerator technology could be applied, the CLIC is expected to be accessible at an energy frontier of around 3000 GeV~\cite{eeLC,CLILC1}.

The production of neutralinos/charginos begins to come into question as a ``discovery channel" of SUSY. Particularly, the production of the neutralinos $\neutralino_{1}$ and $\neutralino_{2}$ at the ILC can yield significant information about the SUSY-breaking mechanism and the nature of the dark matter. The ILC can separate the chargino and neutralino contributions in many cases, providing mass and cross section measurements at the O(1)$\%$ level~\cite{HBaer2}. Particularly, precise predictions for the pair production of neutralinos are significant to derive mass exclusion limits and, should SUSY be realized in nature, to accurately measure the masses and properties of the neutralinos and other sparticles. Motivated by the above considerations, we investigate the center-of-mass energy and the most relevant model parameters dependencies of the neutralino pair production via the process $\gamma\gamma\to \neutralino_{i} \neutralino_{j}$ with complete one-loop Feynman diagrams at the ILC, considering the allowed parameter region in the MSSM. We also analyze the impact of the individual contributions coming from different types of diagrams on the size of the cross section. There have been few papers dedicated to the investigations of this process in the literature as follows. The neutralino pair production via photon-photon fusion in the framework of minimal Super Gravity (mSUGRA) with some typical parameter sets has been previously investigated in the Ref.~\cite{Fei} as a function of neutralino mass and tan$\beta$, and the contribution to the cross section is said to be predominantly from s-channel diagrams in the resonant effect energy region. Considering the helicity nature of the cross sections, a detailed analysis of the process $\gamma\gamma\to \neutralino_{i} \neutralino_{j}$ at a $\gamma\gamma$ linear collider has been carried out in Ref.~\cite{Gounaris} for several benchmark points in the mSUGRA and the gauge mediated supersymmetry breaking (GMSB) models, and this process is emphasized to be very sensitive to the actual values of the various MSSM parameters and details of the neutralino mixings in particular. The production rates of neutralino pairs via photon-photon collisions at the one-loop level including all possible Feynman diagrams in a future photon collider, especially the angular dependence of the cross section, have been calculated in the Ref.~\cite{Nasuf} for four different SUSY benchmark points presented in light of LHC8, and it has been concluded that the contribution to the cross section is mainly from the box-type diagrams. We note that our results are consistent with those obtained in Refs.~\cite{Gounaris,Nasuf}.

Unlike the above-mentioned works, within the present work, the most outstanding feature of our approach is the mechanism in selecting the input parameters. The relevant Lagrangian parameters are recovered as direct analytical
expressions of appropriate physical masses without any restrictions on them in the MSSM. As a matter of fact,
we mainly carry out the algebraically nontrivial inversion in order to obtain Higgsino and gaugino mass parameters.
More explicitly, we can say that using tan$\beta$ and masses of charginos as input parameters, then we get
the other ones being Higgsino/gaugino mass parameters, neutralino masses, and mixing matrix.

The rest of the present work is organized as follows. In Sec. \ref{sec:cros}, we present the corresponding Feynman diagrams and briefly review the standard formulas related to the calculation of cross section for neutralino pair production via photon-photon collisions in the linear colliders. We then provide information on how we have carried out the numerical evaluation. In Sec. \ref{sec:input}, we present some definitions corresponding to our method and input parameters. In Sec. \ref{sec:results}, we give numerical
results and discuss the corresponding SUSY parameter dependences of the cross section for each scenario in detail. Finally,  Sec.~\ref{sec:conclusion} is devoted to conclusions and the highlights of the study.

\section{Calculation of cross section for the neutralino pair production in photon-photon fusion}\label{sec:cros}

\begin{figure*}[!t]
    \begin{center}
\includegraphics[width=\linewidth]{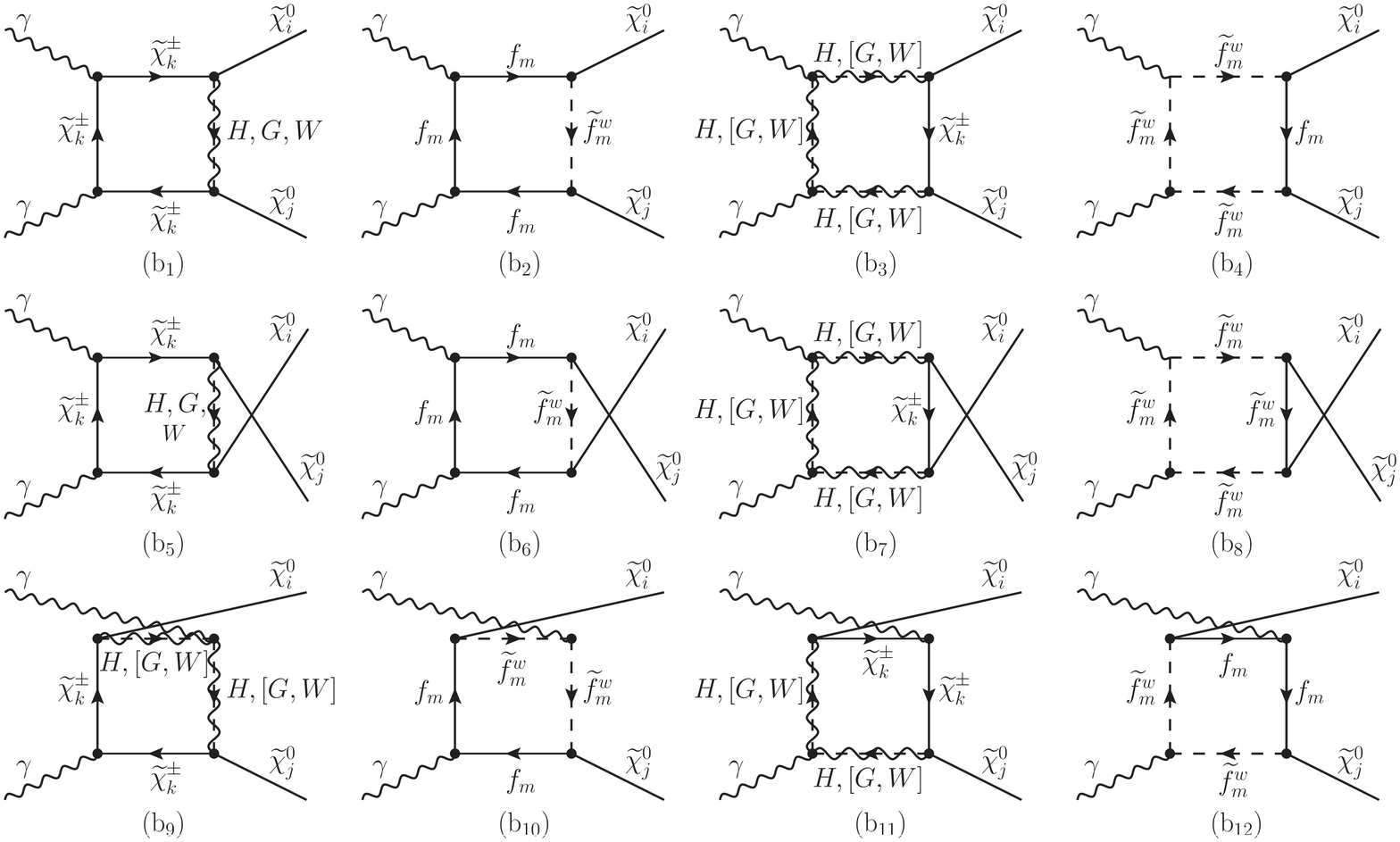}
     \end{center}
\caption{Box-type diagrams contributing to the process $\gamma\gamma\rightarrow\neutralino_{i}\neutralino_{j}$
at one-loop level.}\label{fig:fig1}
\end{figure*}
\begin{figure*}[ht]
    \begin{center}
\includegraphics[width=\linewidth]{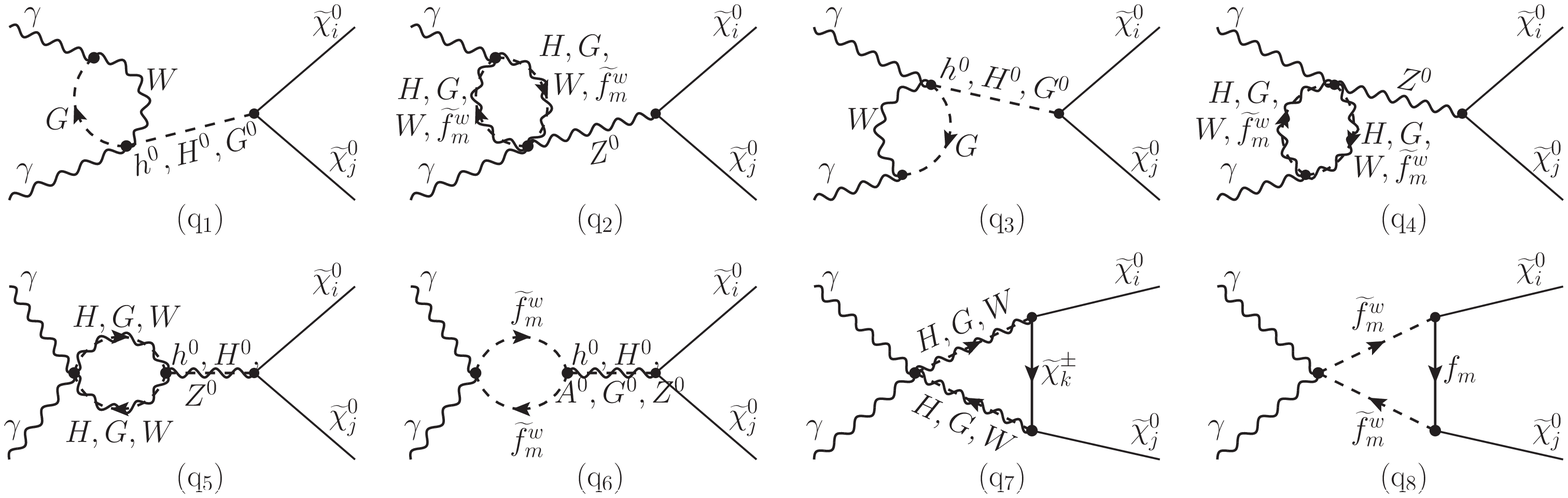}
     \end{center}
\caption{Quartic coupling-type diagrams contributing to the process $\gamma\gamma\rightarrow\neutralino_{i}\neutralino_{j}$
at one-loop level.}\label{fig:fig2}
\end{figure*}
\begin{figure*}[ht]
    \begin{center}
\includegraphics[width=\linewidth]{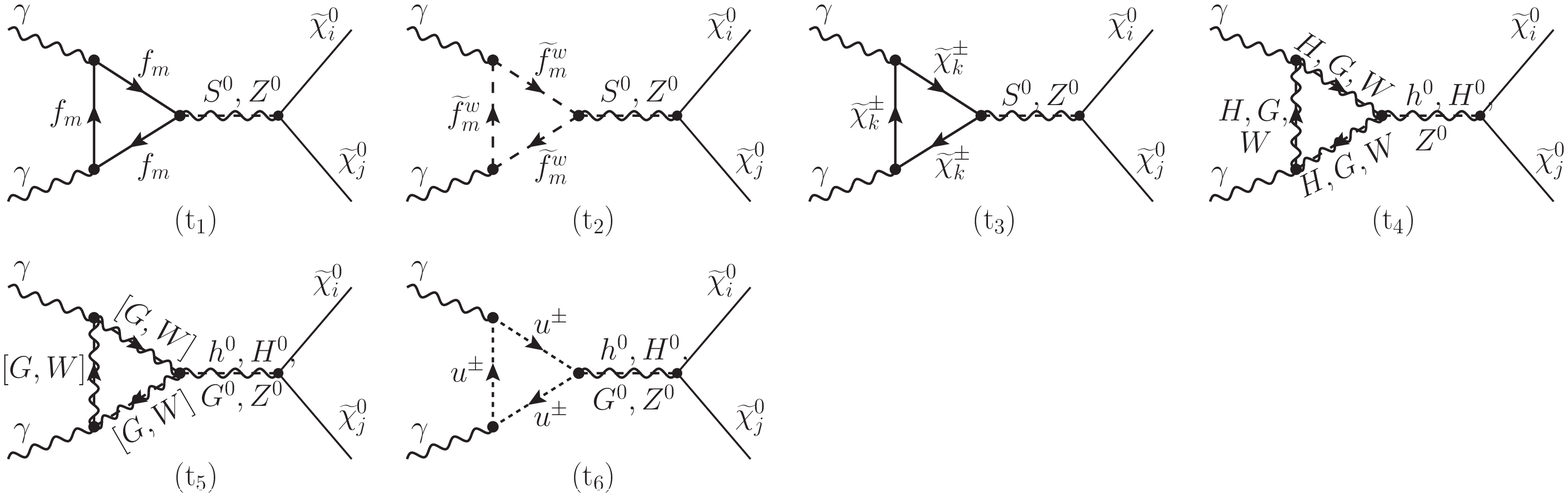}
     \end{center}
\caption{Triangle-type diagrams contributing to the process $\gamma\gamma\rightarrow\neutralino_{i}\neutralino_{j}$
at one-loop level.} \label{fig:fig3}
\end{figure*}
For producing the lighter supersymmetric particles such as neutralinos and charginos, a linear collider would be best suited. The production of a pair of neutralinos in the linear colliders can occur via either electron-positron annihilation or photon-photon collision. In finding the existence of neutralinos, however, photon-photon collision at a linear collider has an advantage over the situation of a linear collider operating in the electron-positron collision mode, where the resonance effects of the intermediate neutral Higgs bosons could be observed only at some specific center-of-mass energies of the collider and the production rate of neutralino pairs would be mostly reduced by the possible s-channel suppression. We consider the process of the neutralino pair production via photon-photon collision at one-loop level, which is denoted by
\begin{equation} \label{eq:gammagammaNN}
\gamma(p_1)\gamma(p_2)\rightarrow\widetilde\chi_{i}^{0}(k_1)\widetilde\chi_{j}^{0}(k_2),
\end{equation}
where, as usual, after each particle we have put the notations for its 4-momenta in parentheses. There is no tree-level amplitude for this subprocess, and it arises for the first time at one-loop level. With the help of the \texttt{FeynArts} (version 3.9) \cite{Feynarts}, we generate a complete set of Feynman diagrams contributing to the subprocess at the one-loop level in the MSSM. These diagrams are displayed in Figs.~\ref{fig:fig1} to~\ref{fig:fig3}. We note that there is also another set of diagrams not explicitly shown in the figures where particles are running counterclockwise in each loop. The
internal particles in diagrams are labelled as follows: the label $S^0$ represents all neutral Higgs/Goldstone bosons $h^0, H^0, A^0, G^0$, and the label $\widetilde{f}^w_m~({f}_m)$ refers to scalar fermions (fermions) $\widetilde{e}^w_m, \widetilde{u}^w_m, \widetilde{d}^w_m~(e_m,u_m,d_m)$. The subscript $m$ and superscript $w$ refer to the generation of (s)quark and the squark mass eigenstates, respectively. The label $k$ running from 1 to 2 represents the type of the charginos. Moreover, $[G,W]$ represents that the loop can consist of all possible combinations of these bosons.

Any one-loop amplitude could be given as a linear sum of triangle, box,
bubble, and tadpole one-loop integrals. In light of this, the one-loop Feynman diagrams contributing to the subprocess \eqref{eq:gammagammaNN} can be divided into three kinds of groups in terms of the one-loop correction type. These are the box-type, triangle-type, and quartic coupling-type diagrams. We have drawn all possible box diagrams, which include the loops of quarks, leptons, squarks and sleptons of three generations, charginos $\chargino_{1,2}$, $W$-boson, charged Higgs boson, and charged Goldstone boson as displayed in Fig.~\ref{fig:fig1}. These diagrams have $t$- and $u$-channel contributions. In Fig.~\ref{fig:fig2}, we show all possible quartic interaction diagrams which consist of bubbles (q$_{1-6}$) connected to the final state through an intermediate Z, or neutral Higgs or Goldstone boson, and triangle loop (q$_{7}$, q$_{8}$) of the $W$-boson, charged Higgs, Goldstone boson, charginos $\chargino_{1,2}$, sfermions, and fermions directly connected to the final state.  In Fig.~\ref{fig:fig3}, we plot all triangle diagrams which include triangle vertices (t$_{1-6}$) connected to the final state through an intermediate Z, or neutral Higgs or Goldstone boson. Feynman diagrams q$_{1-6}$ in Fig.~\ref{fig:fig2} and t$_{1-6}$ in Fig.~\ref{fig:fig3} are referred to as s-channel diagrams. It is obvious that the resonant effects can be observed only in the bubbles-type (q$_{1-6}$) and triangle diagrams (t$_{1-6}$) at some specific center-of-mass energy due to the intermediated neutral Higgs bosons. It should be emphasized that the couplings of the neutralino to particles appearing in diagrams are directly proportional to the elements of the neutralino mixing matrix $N_{ij}$ ($i,j=1,...,4$). This matrix also controls the gaugino/Higgsino fractions of the neutralinos. Therefore, the amplitude of the contribution coming from the different channel diagrams can be dominated by purely gaugino or purely Higgsino or mixed gaugino-Higgsino production.

Since the process $\gamma\gamma\to\widetilde{\chi}_{i}^0\widetilde{\chi}_{j}^0$ in the lowest order has only one-loop contributions, the relevant amplitude could be simply calculated by summing all unrenormalized reducible and irreducible contributions at one-loop level. As a result, we can get the result to be finite and gauge invariant. In calculations, therefore, it is not necessary to take into account the renormalization for ultraviolet divergence which should be canceled automatically. The corresponding Lorentz-invariant matrix element\footnote{We do not present an explicit expression for matrix element since it is too lengthy to include here.} can be given as a sum over contributions coming from box-type, quartic-type, and triangle-type diagrams in the form
\begin{equation}\label{eq:totalM}
{\cal M}= {\cal M}_{box}+ {\cal M}_{quartic}+{\cal M}_{triangle},\
\end{equation}
where there appear a relative sign $(-1)^{\delta_{ij}}$ between the amplitudes of one diagram and its counterpart occurring by interchanging the final states due to Fermi statistics.
Using the total matrix element \eqref{eq:totalM}, the total partonic cross section for the subprocess in unpolarized photon-photon collisions can be calculated by
\begin{equation} \label{eq:totalsigma}
\hat{\sigma}(\hat{s},\gamma\gamma\rightarrow\widetilde\chi_{i}^{0}\widetilde\chi_{j}^{0})=\frac
{1}{16\pi \hat s^{2}}\left(\frac{1}{2}\right)^{\delta_{ij}}\int_{\hat
t^{-}}^{\hat t^{+}}d\hat t \overline{\sum} |{\cal M}|^2,
\end{equation}
where the upper and lower bounds of the integral are defined as
$\hat{t}^\pm=1/2\bigl[(m_i^2+m_j^2-\hat s) \pm\sqrt{(\hat
s-m_i^2-m_j^2)^2-4m_i^2 m_j^2}\bigr]$. The factor $\left(\frac{1}{2}\right)^{\delta_{ij}}$ in the above equation is due to the identical neutralinos in the final state. The bar over the sum refers to the average over initial spins.

The photon-photon collision can be realized at the facility of the International $e^+ e^-$ Linear Collider.   Then, the neutralino pair via photon-photon collision is produced as a subprocess of electron-positron collisions at the ILC. Having computed the partonic cross section with Eq.~\eqref{eq:totalsigma}, the total cross section of the neutralino pair production via photon-photon collisions in electron-positron colliders could be readily obtained by using the formula
\begin{align} \label{eq:total_cross}
&\sigma(s, e^+e^-\rightarrow\gamma\gamma\rightarrow  \widetilde\chi_i^0\widetilde\chi_j^0)=\nonumber\\
&\int_{(m_{\widetilde\chi_i^0}+m_{\widetilde\chi_j^0})/\sqrt{s}}^{x_{max}} dz \frac{dL_{\gamma\gamma}}{dz}~ \hat{\sigma}( \gamma\gamma\rightarrow  \widetilde\chi_i^0\widetilde\chi_j^0;\; \hat{s}=z^2s )\,,
\end{align}
with the photon luminosity
\begin{equation}
\frac{dL_{\gamma\gamma}}{dz}=2z\int_{z^2/x_{max}}^{x_{max}}\frac{dx}{x}F_{\gamma/e}(x)F_{\gamma/e}\left(\frac{z^2}{x}\right)\,,
\end{equation}
where $ \sqrt{\hat{s}}$ and $\sqrt{s}$ are the center-of-mass energies of $\gamma\gamma$ and $e^+e^-$ collisions, respectively, and  $F_{\gamma/e}(x)$  is the photon structure function of electron beam, depending on fraction $x$ of the longitudinal momentum of the electron beam \cite{Telnov}. The quality of the photon spectra is better for higher x. Nevertheless, the high-energy photons for $x > 2(1+\sqrt{2})\approx 4.8$ can vanish via the production of an electron-positron pair in its collision with a following laser photon. That is why the maximum fraction of the photon energy is chosen as $x_{max}=0.83$. One of the best methods of producing intense photon beams is the use of inverse Compton scattering of laser light by the electron beam of a linear accelerator. For the photon structure function, we take the energy spectrum of the initial-state photons provided as Compton backscattered photons off the electron beams \cite{Telnov}.

We take into account all the possible one-loop-level diagrams, which are generated by using \texttt{FeynArts} (version 3.9) \cite{Feynarts}; therefore, our calculations result in a finite result without the need of the renormalization procedure as mentioned above. We perform numerical calculations in the 't Hooft-Feynman gauge where the photon polarization sum is given by $\sum_\lambda \epsilon_\mu^*(k,\lambda)\epsilon_\nu(k,\lambda)=-g_{\mu\nu}$. We carry out the numerical evaluation using the Mathematica packages \texttt{FeynArts} to obtain the relevant amplitudes in \eqref{eq:totalM}, \texttt{FormCalc} (version 9.2) \cite{Hahn} to supply both the analytical results and a complete Fortran code for numerical evaluation of the squared matrix elements, and \texttt{LoopTools} (version 2.13) \cite{loop} to make the evaluation of the necessary one-loop scalar and tensor integrals. In addition, properties of Higgs bosons in the MSSM are computed with the help of the \texttt{FeynHiggs} (version 2.11.3) \cite{FeynHiggs}. In the numerical treatment,
we use the Compton backscattered photons interfaced via the \texttt{CompAZ} code \cite{Compaz} for the photon structure function.

\section{Parameter Settings}\label{sec:input}
In this section, we briefly give some details regarding our method and input parameters. During our numerical evaluations, we have considered the assumptions and approaches in our previous works \cite{Demirci3, Demirci1, Demirci2}
for the gaugino/Higgsino sector. The soft SUSY-breaking gaugino mass parameters $M_1$
related to the $U(1)$, $M_2$ related to the $SU(2)$, and
the Higgsino mass parameter $\mu$ could be taken to be real and positive. Furthermore, the gaugino mass parameter $M_1$ is commonly
fixed by way of the relation $M_1=\frac{5}{3}M_2
\tan^2\theta_W\simeq 0.5 M_2$. The parameters $M_2$ and $\mu$ are analytically derived by taking the suitable difference and sum of the chargino masses as shown in Eqs. (A13) and (A14)
in Ref. \cite{Demirci2}.
Consequently, there appear three different cases in the selection of the gaugino/Higgsino mass
parameters $M_2$ and $\mu$. These are the Higgsino-like, gauginolike, and mixture case, separately. To examine the phenomenological consequences in a most general approach, we consider the
three possible scenarios. For more details on the definition of these scenarios, we refer to Ref.~\cite{Demirci2}. We set the chargino masses as
\begin{equation} \label{eq:mc}
m_{\chargino_{1}}=214.31\gev,~m_{\chargino_{2}}= 671.56\gev
\end{equation}
for both Higgsino-like and gauginolike scenarios and
\begin{equation} \label{eq:mc2}
m_{\chargino_{1}}= 271.68\gev,~m_{\chargino_{2}}= 396.13\gev
\end{equation}
for the mixture-case scenario. Then, for given any $\tan\beta$, the values of parameters $M_2$ and $\mu$ in each scenario are calculated by using values given in Eqs.~\eqref{eq:mc} and~\eqref{eq:mc2}. Subsequently, neutralino masses for each scenario are computed, depending on the values of $\mu$ and $M_2$.
Furthermore, in view of the constraints derived on SUSY parameters from available experimental data, especially the LHC results from the 7 and 8 TeV data for the direct SUSY research \cite{sqgl_ATLAS1,sqgl_ATLAS2,sqgl_CMS1,sqgl_CMS2}, we set the soft SUSY-breaking parameters for the entries of mass matrices in the slepton and squark sector to be equal as $M_{\widetilde{L},\widetilde{E}}=1\tev$ and $M_{\widetilde{Q},\widetilde{U},\widetilde{D}}=2\tev$, respectively, and fix $\tanb=10,~m_{A^{0}} = 1000\gev$, and $A_{t,b,\tau}=2700\gev$ where $m_{A^{0}}$ is the mass of the pseudoscalar Higgs boson and $A_{t,b,\tau}$ are the trilinear couplings. Moreover, we take the input parameters for the SM, $m_W= 80.385\gev$, $m_Z= 91.1876\gev$, $\alpha^{-1}= 137.036$, and $\alpha(m_Z^2)^{-1}= 127.934$ \cite{PDG}, and ignore the masses of the light quarks.

We give a list of the Higgsino/gaugino mass parameters and the masses of the neutral/charged Higgs bosons and neutralinos, which have been obtained according to the above default parameters, for each scenario in Table~\ref{table1}. Here, the masses of Higgs bosons have been calculated to two-loop accuracy by using the \texttt{FeynHiggs}. Note that one can always adjust the squark masses and the trilinear couplings in order to ensure a correct Higgs boson mass. The mass of the lightest Higgs boson is obtained to be consistent with the SM predictions.
\begin{table*}[htb]
\caption{The Higgsino/gaugino mass parameters, masses of Higgs bosons, and neutralinos for each scenario, where all masses are given in GeV.}\label{table1}
\begin{ruledtabular}
\begin{tabular}{lR[.][.]{3}{2}R[.][.]{3}{2}R[.][.]{3}{2}R[.][.]{3}{2}R[.][.]{4}{2}R[.][.]{4}{2}R[.][.]{3}{2}R[.][.]{3}{2}R[.][.]{3}{2}R[.][.]{3}{2}}
 &\multicolumn{1}{c}{$M_{2}$}&\multicolumn{1}{c}{$\mu$}&\multicolumn{1}{c}{$M_{1}$}
&\multicolumn{1}{c}{$m_{h^0}$}&\multicolumn{1}{c}{$m_{H^0}$}&\multicolumn{1}{c}{$m_{H^\pm}$}&\multicolumn{1}{c}{$m_{\neutralino_{1}}$}& \multicolumn{1}{c}{$m_{\neutralino_{2}}$}&\multicolumn{1}{c}{$m_{\neutralino_{3}}$}&\multicolumn{1}{c}{$m_{\neutralino_{4}}$}\\
\hline
Higgsino-like&660.00&220.00&315.51&125.93&1000.08&1003.34&202.23&224.37&325.58&671.59\\
Gauginolike& 220&660.00& 105.17&124.92&1000.08&1003.34& 103.95&214.36&663.92&670.61\\
Mixture case&330.00&330.00&157.76&125.52&1000.10&1003.37&152.33&274.03&335.47&396.61\\
\end{tabular}
\end{ruledtabular}
\end{table*}
\begin{table*}[ht]
\caption{The gaugino/Higgsino fractions of the lightest and the next-to-lightest neutralinos for each scenario. Here, note that the sum of these fractions is equal to 1 for each type of neutralino.}\label{table2}
\begin{ruledtabular}
\begin{tabular}{lR[.][.]{3}{3}R[.][.]{3}{3}R[.][.]{3}{3}R[.][.]{3}{3}}
~~&\multicolumn{2}{c}{$\widetilde{\chi}_{1}^{0}$}&\multicolumn{2}{c}{$\widetilde{\chi}_{2}^{0}$}\\ \cline{2-5}
 &\multicolumn{1}{c}{$|{N_{11}}|^2+|N_{12}|^2$}&\multicolumn{1}{c}{$|{N_{13}}|^2+|N_{14}|^2$}&\multicolumn{1}{c}{$|N_{21}|^2+|N_{22}|^2$}&\multicolumn{1}{c}{$|N_{23}|^2+|N_{24}|^2$}\\
\hline
Higgsino-like&0.090&0.910&0.006&0.994\\
Gauginolike&0.995&0.005&0.978&0.022\\
Mixture case&0.956&0.044&0.536&0.464\\
\end{tabular}
\end{ruledtabular}
\end{table*}

It is well known that in terms of the elements of the mixing matrix $N$, neutralinos are constructed by the following mixture of gaugino and Higgsino components:
\begin{equation} \label{eq:NN}
\neutralino_{i}=N_{i1}\widetilde{B}+N_{i2}\widetilde{W}^3+N_{i3}\widetilde{H}^0_1+N_{i4}\widetilde{H}^0_2
\end{equation}
It is obvious from Eq.~\eqref{eq:NN} that the quantities $|N_{i1}|^2+|N_{i2}|^2$ and $|N_{i3}|^2+|N_{i4}|^2$ represent to the rate of the gaugino component and Higgsino component which are contained in a neutralino $\neutralino_{i}$, respectively. These quantities are often referred to as the gaugino fraction and Higgsino fraction of the neutralino and  can be used to easily determine the nature (character) of the neutralino. We can point out that both the lightest neutralino $\neutralino_{1}$ and the next-to-lightest neutralino $\neutralino_{2}$ have a dominant Higgsino component in the Higgsino-like scenario and a dominant gaugino (bino or wino) component in the gauginolike scenario as seen from Table~\ref{table2}. Moreover, in the mixture-case scenario, the neutralino $\neutralino_{1}$ has a relatively large gaugino component, while the neutralino $\neutralino_{2}$ consists of gaugino and Higgsino components being almost equal to each other\footnote{Additionally, unlike the scenario considered in this study, there appear other hierarchies between Higgsino and gaugino mass parameters  $\mu$, $M_1, M_2$. For instance, $M_1<\mu<M_2$; if this hierarchy were considered,  a situation in which the neutralino $\neutralino_{1}$ is mostly gaugino and the neutralino $\neutralino_{2}$ is mostly a Higgsino would be easily gotten.}. In light of these scenarios, thus, one can easily determine what components of the produced neutralinos are more dominant when the enhancement of cross section is large enough to be detectable at the ILC.

\section{Numerical results and discussion}\label{sec:results}

In this section, we discuss in detail the numerical predictions of the neutralino pair production via photon-photon fusion at the ILC, focusing on gaugino/Higgsino fractions of the neutralinos. For representative parameter points of each scenario given in Table~\ref{table1}, we have carried out numerical evaluation of the total cross sections of process $e^{+} e^{-} \to\gamma\gamma\to\widetilde{\chi}_{i}^0\widetilde{\chi}_{j}^0$ as a function of the center-of-mass energy and the tan$\beta$ and subprocess $\gamma\gamma\to\widetilde{\chi}_{i}^0\widetilde{\chi}_{j}^0$ on the $M_2$-$\mu$ mass plane for $i=j=1,2$\footnote{We note that the final state $\widetilde{\chi}_{1}^0\widetilde{\chi}_{1}^0$ is invisible in an R-parity
invariant theory where the lightest neutralino $\widetilde{\chi}_{1}^0$ is the LSP. For completeness, we have also provided a calculation of the production $\widetilde{\chi}_{1}^0\widetilde{\chi}_{1}^0$.  However, it could be investigated indirectly by $\gamma$ tagging in the final state $\gamma \widetilde{\chi}_{1}^0\widetilde{\chi}_{1}^0$ (see Ref. \cite{Pandita}), which can be observed with a rather high accuracy at a linear collider.}. To examine how the contributions of different types of diagrams have effects on the total cross section, we have also computed the individual contributions stemming from each type of diagram as a function of the center-of-mass energy. Note that, during our calculations, we have numerically checked the cancellation of divergences appearing in the loop contributions, giving, as expected, a finite result without the need of the renormalization procedure. We use the following shorthand: HL is Higgsino-like, GL is gauginolike, and MC is mixture-case.

\begin{figure*}[thp]
    \begin{center}
\includegraphics[scale=0.8]{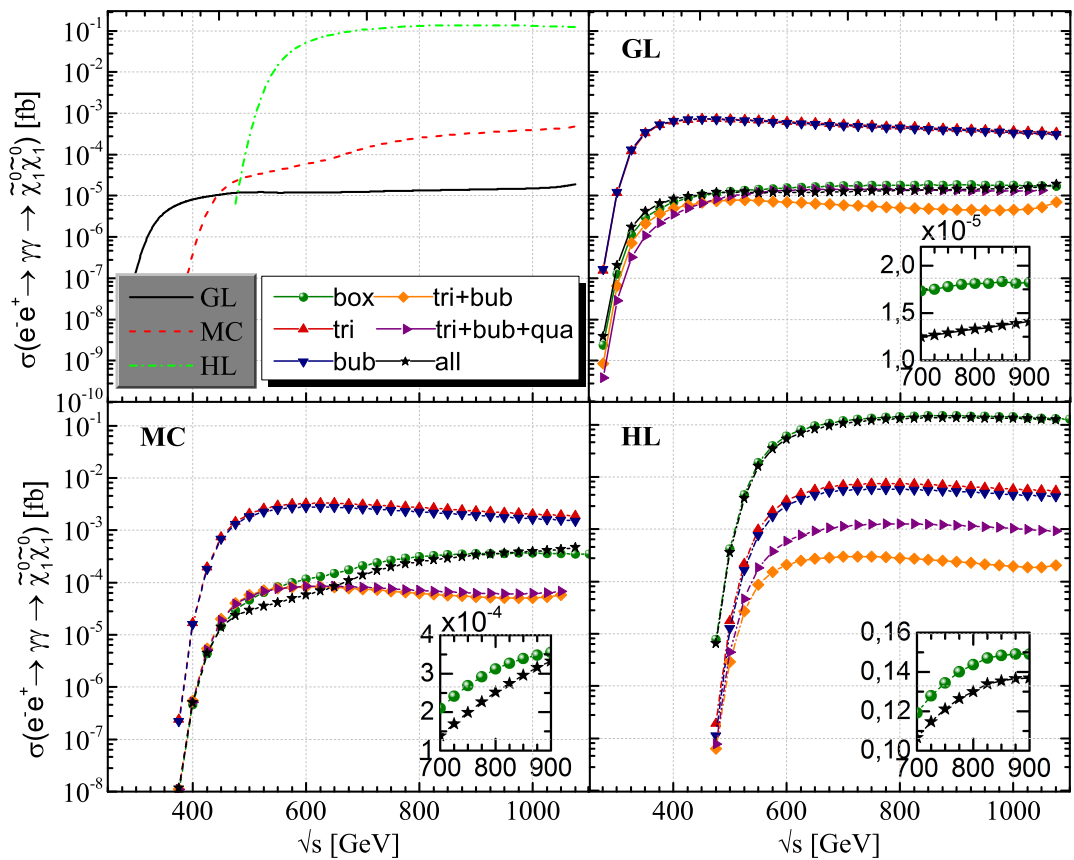}
     \end{center}
\caption{(color online). Total  cross section (the upper left panel) and the individual contributions coming from different types of diagrams (the other panels) for the process $e^{+} e^{-} \rightarrow\gamma\gamma\rightarrow\widetilde{\chi}_{1}^0\widetilde{\chi}_{1}^0$ vs the center-of-mass energy of the $e^{+} e^{-}$ collider. The inserts show the contributions of the box-type and all diagrams with varying center-of-mass energy from 700 to 900 GeV.} \label{fig:fig4}
\end{figure*}

\begin{figure*}[hpt]
    \begin{center}
\includegraphics[scale=0.8]{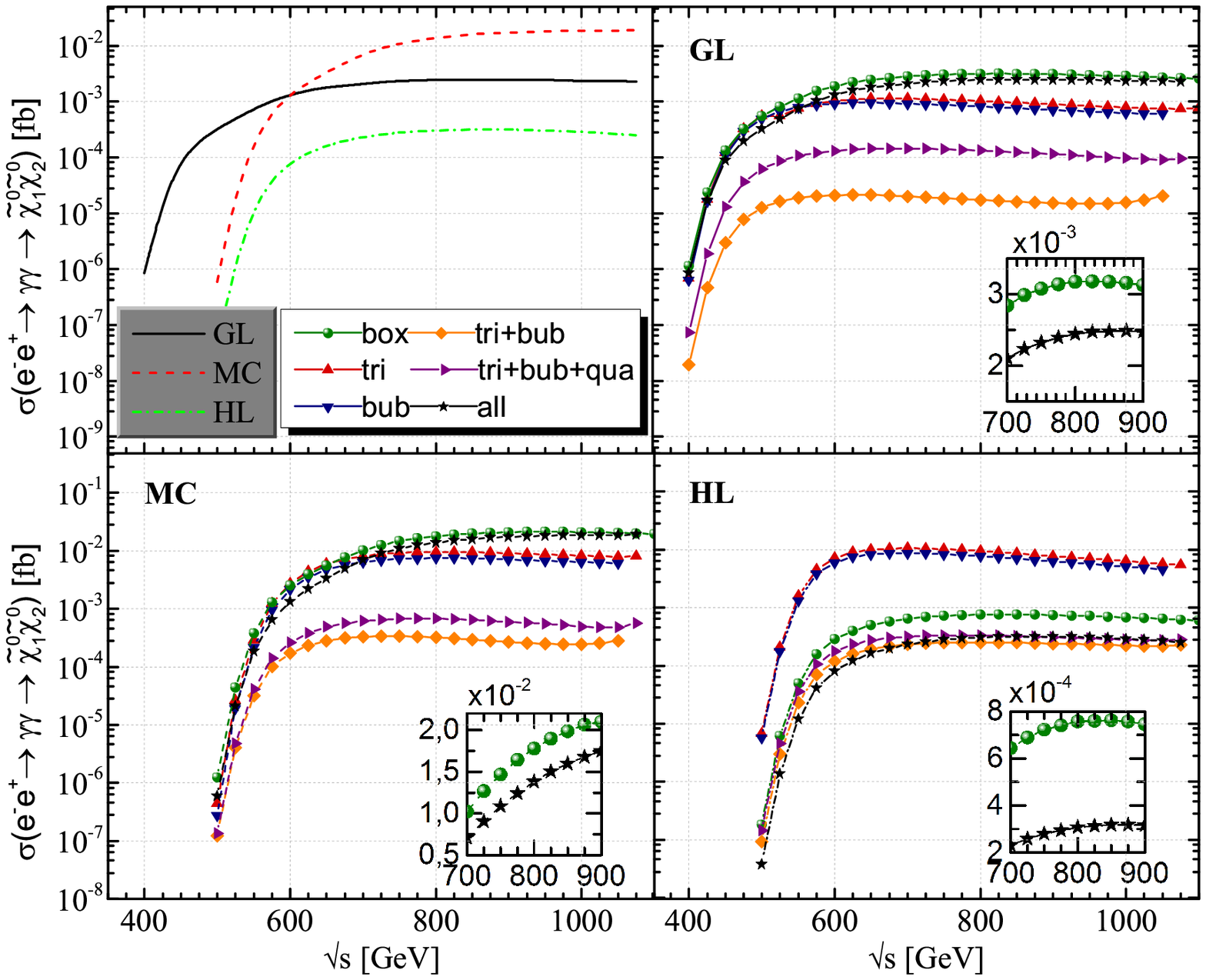}
     \end{center}
\caption{(color online). Total  cross section (the upper left panel) and the individual contributions coming from different types of diagrams (the other panels) for the process $e^{+} e^{-} \rightarrow\gamma\gamma\rightarrow\widetilde{\chi}_{1}^0\widetilde{\chi}_{2}^0$ vs the center-of-mass energy of the $e^{+} e^{-}$ collider. The inserts show the contributions of the box-type and all diagrams with varying center-of-mass energy from 700 to 900 GeV.} \label{fig:fig5}
\end{figure*}

\begin{figure*}[hpt]
    \begin{center}
\includegraphics[scale=0.8]{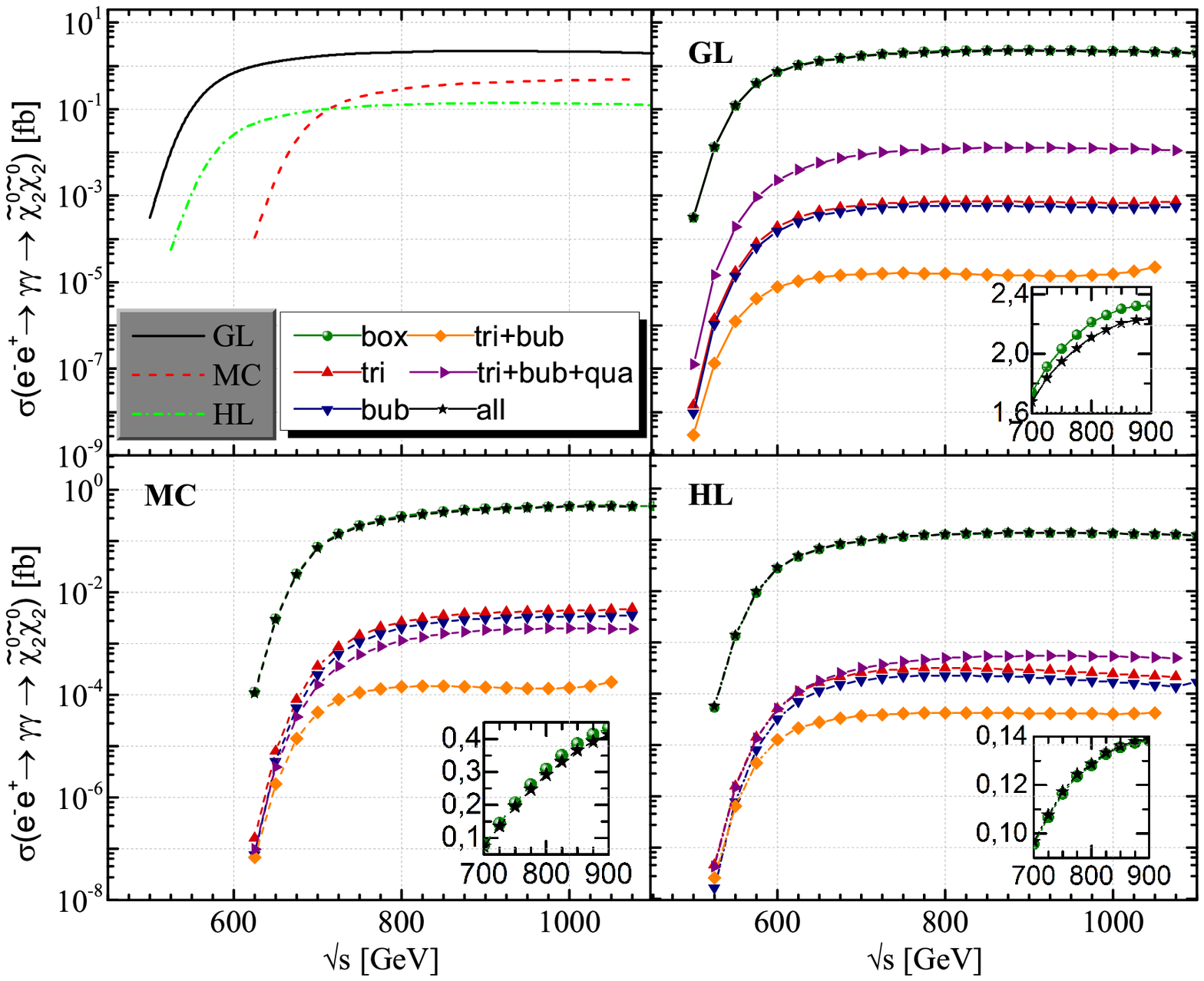}
     \end{center}
\caption{(color online). Total  cross section (the upper left panel) and the individual contributions coming from different types of diagrams (the other panels) for the process $e^{+} e^{-} \rightarrow\gamma\gamma\rightarrow\widetilde{\chi}_{2}^0\widetilde{\chi}_{2}^0$ vs the center-of-mass energy of the $e^{+} e^{-}$ collider. The inserts show the contributions of the box-type and all diagrams with varying center-of-mass energy from 700 to 900 GeV.} \label{fig:fig6}
\end{figure*}

In Figs.~\ref{fig:fig4} through \ref{fig:fig6}, we show the dependence of the total cross sections and the individual contributions coming from different types of diagrams on the center-of-mass energy ranging from 200 to 1100 GeV for the neutralino pair production via photon-photon fusion in the parent $e^{+} e^{-}$ collider. In these figures, the abbreviations  ``box'', ``bub'',``qua'',``tri'' and ``all'' correspond to the contributions of box-type (diagrams b$_{1-12}$ in Fig.~\ref{fig:fig1}), bubble-type (diagrams q$_{1-6}$ in Fig.~\ref{fig:fig2}), quartic-type (diagrams q$_{7}$ and q$_{8}$ in Fig.~\ref{fig:fig2}), triangle-type (diagrams t$_{1-6}$ in Fig.~\ref{fig:fig3}) diagrams, and the sum of all the considered Feynman diagrams, respectively. Also, the ``tri+bub'' is referred to as the contribution arising from interference between triangle-type and bubble-type diagrams.
These plots reveal that the total cross sections are getting bigger very quickly in a small interval of center-of-mass energy at which the production is kinematically possible for the first time and then continue to increase slowly and remain almost stable with increments of the center-of-mass energy for each scenario. When the center-of-mass energy of the $e^{+} e^{-}$ collider is in the vicinity of $1100\gev$, the total cross sections remarkably increase (i.e., the resonance effects of the intermediate neutral Higgs bosons for $\sqrt{s}\gtrsim1100$ GeV, which are not explicitly shown in the figures, appear) because the maximum center-of-mass energy of incoming photons is close to $\sqrt{\hat{s}}\sim 0.83\sqrt{s}<m_{H^0,A^0}$. One can say that there appear energy-dependent structures of the total cross section at the center-of-mass energy being close to the sum of the masses of intermediate particles as well as resonance effects due to diagrams involving the intermediate neutral Higgs boson $H^0$ or $A^0$. Moreover, it is clearly seen that the process $e^{+}e^{-}\to\gamma\gamma\to\widetilde{\chi}_{i}^0\widetilde{\chi}_{j}^0$ is mostly dominated by the Higgsino-like scenario in the case of $i=1,j=1$; the micture-case scenario in the case of $i=1,j=2$; and the gauginolike scenario in the case of $i=2,j=2$. Further, a detailed analysis
indicates that the largest values of the cross section are obtained when the lightest neutralino $\widetilde{\chi}_{1}^0$ has a dominant Higgsino component and/or the next-to-lightest neutralino $\widetilde{\chi}_{2}^0$ has a dominant gaugino component. The sizes of the total cross sections are at a visible level of $10^{-1}$ fb for $e^{+}e^{-}\to\gamma\gamma\to\widetilde{\chi}_{1}^0\widetilde{\chi}_{1}^0$ in the HL scenario, $10^{-2}$ fb for $e^{+}e^{-}\to\gamma\gamma\to\widetilde{\chi}_{1}^0\widetilde{\chi}_{2}^0$ in the MC scenario, and $10^{1}$ fb for $e^{+}e^{-}\to\gamma\gamma\to\widetilde{\chi}_{2}^0\widetilde{\chi}_{2}^0$ in the GL scenario. In particular, the total cross section of the production of $\widetilde{\chi}_{2}^0\widetilde{\chi}_{2}^0$ reaches a value of 2.23 fb at $\sqrt{s}= 900$ GeV and may be more observable than others at the ILC for the GL scenario. Furthermore, the cross sections are sorted in descending order according to our scenarios as $\sigma$(HL)$>\sigma$(MC)$>\sigma$(GL) for $i=1,j=1$; $\sigma$(MC)$>\sigma$(GL)$>\sigma$(HL) for $i=1,j=2$; and $\sigma$(GL)$>\sigma$(MC)$>\sigma$(HL) for $i=2,j=2$. We remark that the basic size of the total cross sections differ by few orders of magnitude, depending on the gaugino/Higgsino fractions of the produced neutralinos.

Let us now analyze the effects of individual contributions stemming from different types of diagrams, which are shown in the upper right (for the GL scenario) and the lower panels (for the MC and HL scenarios) of Figs.~\ref{fig:fig4}-\ref{fig:fig6}. Note that for $\sqrt{s}> 1050$ GeV some data points of individual contributions are not marked in these figures because these are becoming very large due to the resonance effects. We clearly see that the dominant individual contribution can change according to the scenarios. This means that it depends on gaugino/Higgsino fractions of the neutralinos. The contributions coming from the box-type diagrams for the HL scenario and both bubble- and triangle-type diagrams (i.e., the s-channel diagrams) for the GL and MC scenarios have a greater effect on the production cross section of $\widetilde{\chi}_{1}^0\widetilde{\chi}_{1}^0$. For the production of $\widetilde{\chi}_{1}^0\widetilde{\chi}_{2}^0$,  the both types of s-channel diagrams in the HL scenario and the box-type diagrams in the GL and MC scenarios give a much larger contribution to the total cross section than the others. Finally, for the production of $\widetilde{\chi}_{2}^0\widetilde{\chi}_{2}^0$, the box-type diagrams have the largest contribution in all of the considered scenarios.

However, it should be emphasized that the contributions of triangle-type and bubble-type diagrams are nearly equal to each other, and the interference between them (tri+bub) always gives a much smaller contribution than each of them (by 1-2 orders of magnitude) for all cases since they almost cancel the contributions of each other. As a matter of fact, they make a destructive interference. Furthermore, the contribution of box-type diagrams is greater than that of this interference. Although the quartic-type diagrams make a positive contribution to the cross section, the sum of triangle-, bubble-, and quartic-type diagrams (tri+bub+qua) has still a low contribution compared to the box-type diagrams. Taking into account all of these, it is worth noting that the box-type diagrams make the main contribution to the total cross section for each scenario. Namely, the box-type diagrams have a significant impact on production rates.

\begin{table*}[!ht]%
\caption{Total cross sections and contributions of box-type, triangle-type, and bubble-type diagrams to the total cross section of the process $e^{+}e^{-}\to\gamma\gamma\to\widetilde{\chi}_{i}^{0}\widetilde{\chi}_{j}^{0}$ ($i,j=1,2$) at the collision energies of electron-positron $\sqrt s$ being  600 and 1000 GeV  for all three scenarios. Here, three dots indicate that production of a neutralino pair is not kinematically accessible for a given energy.}\label{table3}
\resizebox{\textwidth}{!}{\begin{tabular}{lccccccccccccc}
 \hline \hline
~~&~&\multicolumn{4}{c}{$\sigma(e^{+}e^{-}\to\gamma\gamma\to\widetilde{\chi}_{1}^{0}\widetilde{\chi}_{1}^{0})$~(fb)}&\multicolumn{4}{c}{$\sigma(e^{+}e^{-}\to\gamma\gamma\to\widetilde{\chi}_{1}^{0}\widetilde{\chi}_{2}^{0})$~(fb)}&\multicolumn{4}{c}{$\sigma(e^{+}e^{-}\to\gamma\gamma\to\widetilde{\chi}_{2}^{0}\widetilde{\chi}_{2}^{0})$~(fb)}\\ \cline{3-14}
S&$\sqrt{s}$~(GeV)&Box&Tri&Bub&All&Box&Tri&Bub&All&Box&Tri&Bub&All\\
 \hline
\multirow{2}*{HL}&~600&0.061&0.004&0.003&0.053&2.9$\cdot10^{-4}$&0.007&0.006&8.1$\cdot10^{-5}$&0.028&5.1$\cdot10^{-4}$&3.3$\cdot10^{-5}$&0.029\\
                        &1000&0.142& 0.006&0.005&0.132&6.8$\cdot10^{-4}$&0.006&0.005&2.9$\cdot10^{-4}$&0.134&2.5$\cdot10^{-4}$&1.7$\cdot10^{-4}$&0.135\\
\noalign{\smallskip}
\multirow{2}*{GL}       &~600&1.6$\cdot10^{-5}$&6.2$\cdot10^{-4}$&5.8$\cdot10^{-4}$&1.2$\cdot10^{-5}$&0.002&0.001&9.3$\cdot10^{-4}$&0.001&0.735&1.9$\cdot10^{-4}$&1.5$\cdot10^{-4}$&0.716\\
                        &1000&1.8$\cdot10^{-5}$& 3.7$\cdot10^{-4}$&3.4$\cdot10^{-4}$&1.5$\cdot10^{-5}$& 0.003&0.001&6.2$\cdot10^{-4}$&0.002&2.236&6.8$\cdot10^{-4}$&5.3$\cdot10^{-4}$&2.154\\
\noalign{\smallskip}
\multirow{2}*{MC} &~600&1.2$\cdot10^{-4}$&0.003&0.003&5.8$\cdot10^{-5}$&0.003&0.003&0.002&0.001&$\cdot\cdot\cdot$&$\cdot\cdot\cdot$&$\cdot\cdot\cdot$&$\cdot\cdot\cdot$\\
                            &1000&3.6$\cdot10^{-4}$& 0.002&0.002&3.9$\cdot10^{-4}$& 0.021&0.008&0.006&0.019&0.487&0.004&0.003&0.469\\
 \hline  \hline
\end{tabular}}
\end{table*}
With a view to make easy precise comparisons with the experimental results, we list a numerical survey over our scenarios for ILC center-of-mass energies of 600 and 1000 GeV in Table~\ref{table3}. It can be easily seen from the above analysis and this table that the total cross section of the process $e^{+}e^{-}\rightarrow\gamma\gamma\to\widetilde{\chi}_{2}^{0}\widetilde{\chi}_{2}^{0}$ in the GL scenario is larger than others. Therefore, this process can be said to be the most dominant in the production of neutralino pairs via photon-photon fusion.

We should also note that, since the masses of the charged Higgs bosons and sfermion are fixed at the TeV scale, the contribution of box-type diagrams  b$_{1}$, b$_{3}$, b$_{5}$, b$_{7}$, b$_{9}$, and b$_{11}$ in Fig.~\ref{fig:fig1} with a W boson in the loop ends up dominating over all the other ones.  About $80\%$ of the sum contribution of box-type diagrams comes from these diagrams. Thus, it is also possible to interpret the contribution of box-type diagrams in terms of a single coupling (the neutralino-chargino-W coupling) and of the chargino masses running in the loop.  The neutralino-chargino-W coupling is proportional to elements of the neutralino mixing matrix $N_{i2}$, $N_{i3}$, and $N_{i4}$ [see Eq.~(C83) of Ref.~\cite{Haber} for corresponding coupling], namely, the wino and Higgsino components of neutralino $\widetilde{\chi}_{i}^{0}$. The cross section of the process $e^{+}e^{-}\rightarrow\gamma\gamma\to\widetilde{\chi}_{1}^{0}\widetilde{\chi}_{1}^{0}$ in the GL scenario is very suppressed (see  Fig.~\ref{fig:fig4}) since in that scenario the lightest neutralino  $\widetilde{\chi}_{1}^{0}$ is mostly a bino (due to the grand unified theory relation between $M_1\simeq 0.5 M_2$) which does not enter in such a coupling (where, indeed, the bino component $N_{i1}$ does not appear). Furthermore, the cross section of the process $e^{+}e^{-}\rightarrow\gamma\gamma\to\widetilde{\chi}_{2}^{0}\widetilde{\chi}_{2}^{0}$ is mostly dominated in the GL scenario (see  Fig.~\ref{fig:fig6}) since the next-to-lightest neutralino  $\widetilde{\chi}_{2}^{0}$ in this scenario is mostly a wino (the wino component $N_{i2}$), which enters in such a coupling.

\begin{figure}[bpt]
    \begin{center}
\includegraphics[scale=0.47]{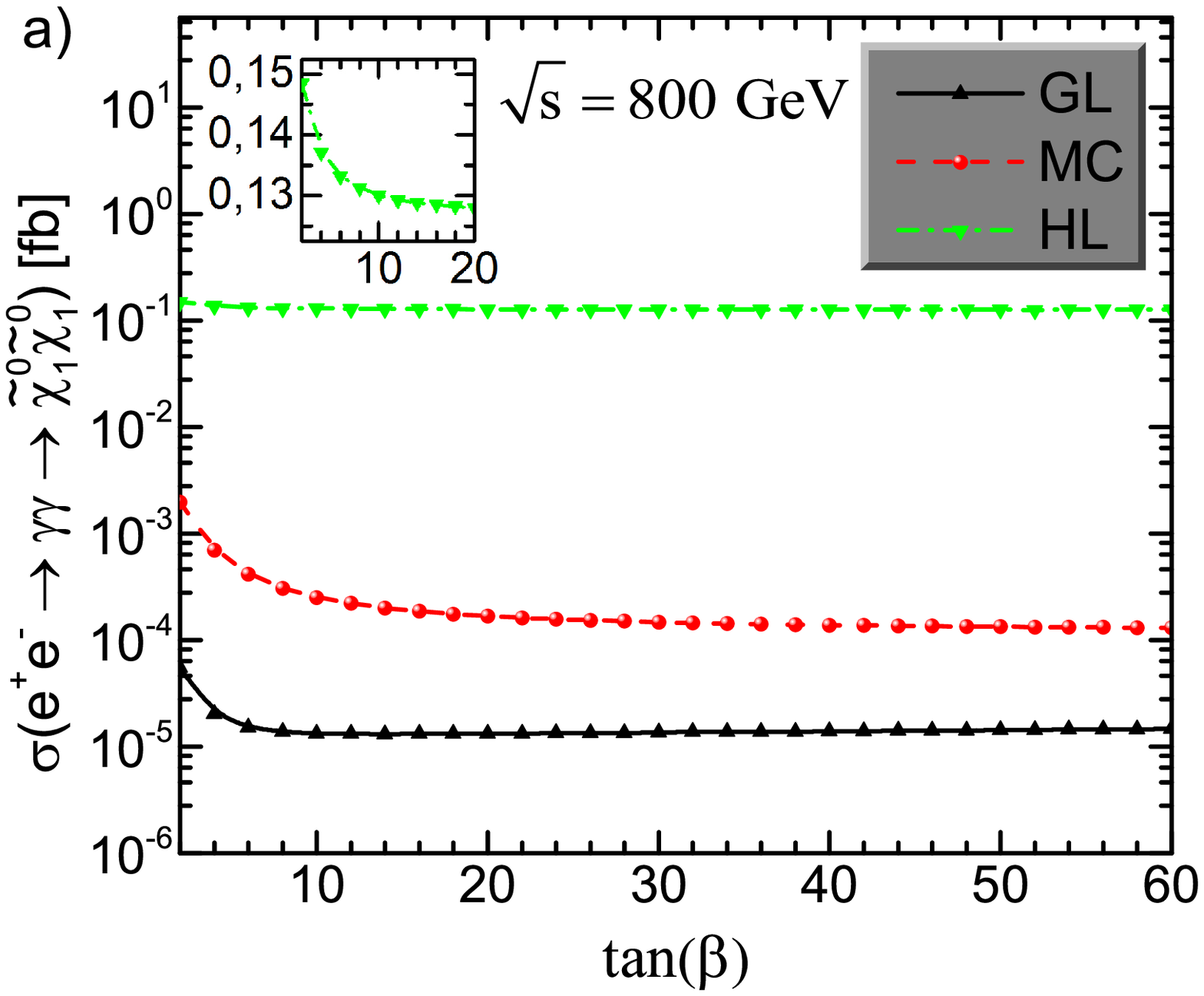}
\includegraphics[scale=0.47]{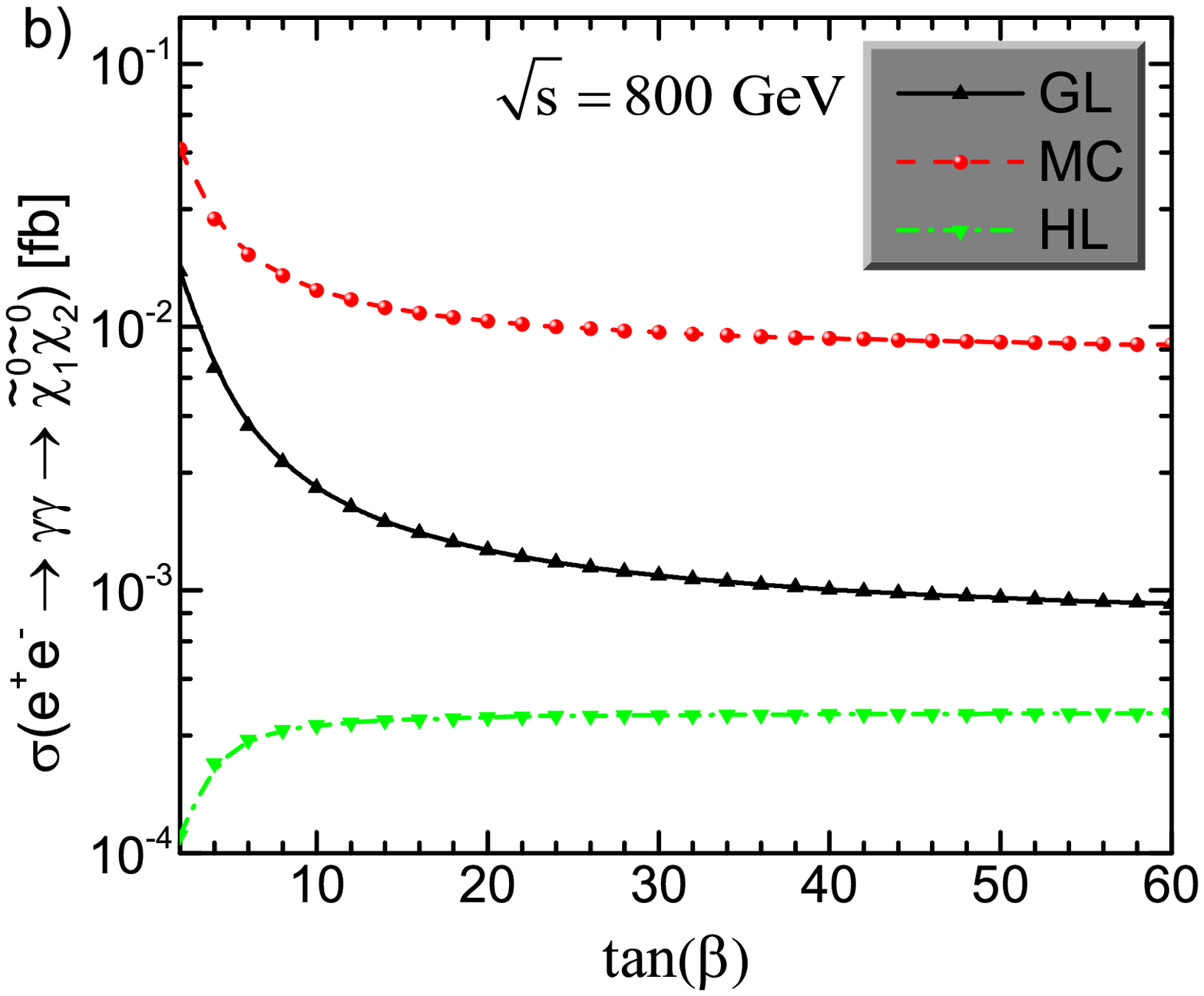}
\includegraphics[scale=0.47]{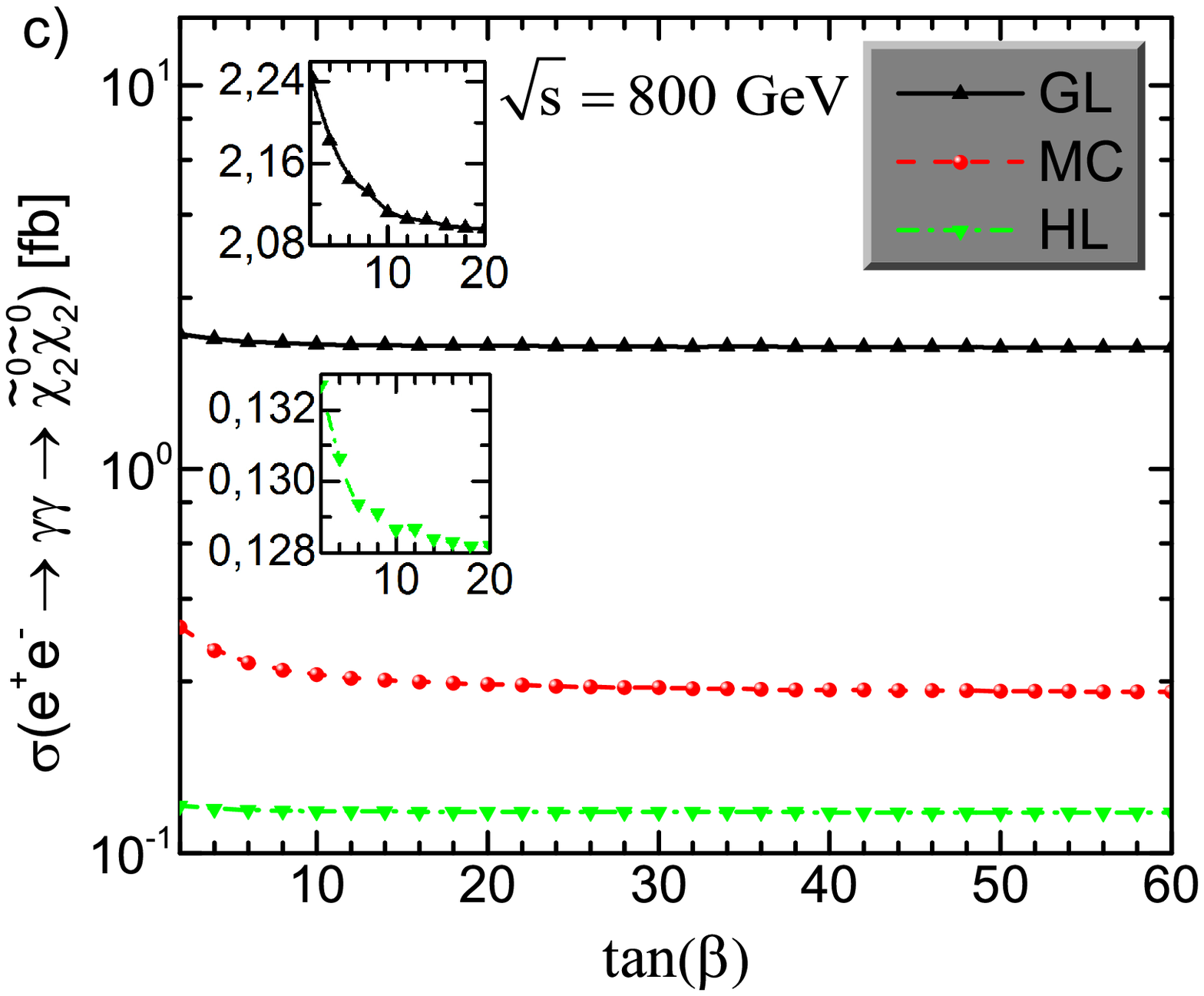}
     \end{center}
\caption{(color online). Total cross sections of the
process $e^{+} e^{-} \rightarrow\gamma\gamma\rightarrow\widetilde{\chi}_{i}^0\widetilde{\chi}_{j}^0$ for (a) $i=j=1$, (b) $i=1,j=2$ and (c) $i=j=2$ depending on the tan$\beta$ at $e^{+} e^{-}$ colliding energy $\sqrt{s}=800$ GeV. Here, we use the inserts to see more clearly the behavior of the cross section in small values of tan$\beta$.} \label{fig:fig7}
\end{figure}
We exhibit the dependence of the total cross section on the tan$\beta$ varied in the range from 2 to 60 at $\sqrt{s}=$ 800 GeV for each scenario in Figs.~\ref{fig:fig7}{\color{red}(a)}-\ref{fig:fig7}{\color{red}(c)}. These figures demonstrate the same dominant scenarios as the ones in the center-of-mass energy dependence of the cross section. The total cross section decreases with increments of tan$\beta$ in most cases. However, for large values of tan$\beta$, it is not very sensitive with respect to the variation of tan$\beta$. That is because the neutralinos masses and mixing matrix are nearly independent of the tan$\beta$, and the dependence of cross section on tan$\beta$ is mostly due to  the masses of Higgs bosons and Yukawa couplings being a function of tan$\beta$. Figure~\ref{fig:fig7}{\color{red}(a)} displays that the cross section of the process $e^{+}e^{-}\to\gamma\gamma \to \widetilde{\chi}_{1}^{0}\widetilde{\chi}_{1}^{0}$ decreases from 0.15 to 0.13 fb in the HL scenario, 1.98$\times10^{-3}$ to 2.52$\times10^{-4}$ fb in the MC scenario, and 5.52$\times10^{-5}$ to 1.33$\times10^{-4}$ fb in the GL scenario when the tan$\beta$ runs from 2 to 10. Figure~\ref{fig:fig7}{\color{red}(b)} shows that the cross section of the process $e^{+}e^{-}\rightarrow\gamma\gamma \to \widetilde{\chi}_{1}^{0}\widetilde{\chi}_{2}^{0}$ varies from 1.16$\times10^{-4}$ to 3.06$\times10^{-4}$ fb in the HL scenario, 0.048 to 0.014 fb in the MC scenario, and 0.016 to 0.002 fb in the GL scenario when the tan$\beta$ goes up from 2 to 10. Figure~\ref{fig:fig7}{\color{red}(c)} presents that the cross section of the process $e^{+}e^{-}\to\gamma\gamma\to\widetilde{\chi}_{2}^{0}\widetilde{\chi}_{2}^{0}$ decreases from 0.13 to 0.12 fb in the HL scenario, 0.39 to 0.29 fb in the MC scenario, and 2.25 to 2.11 fb in the GL scenario when the tan$\beta$ runs from 2 to 10. Also, it is obvious from these figures that the cross section of the process $e^{+}e^{-}\to\gamma\gamma \to\widetilde{\chi}_{1}^{0}\widetilde{\chi}_{1}^{0}$ [shown in Fig.~\ref{fig:fig7}{\color{red}(a)}] in the HL scenario is about 3 and 4 orders of magnitude larger than ones in the MC and GL scenarios, respectively. Furthermore, the cross section of the process $e^{+}e^{-}\to\gamma\gamma\to\widetilde{\chi}_{1}^{0}\widetilde{\chi}_{2}^{0}$ [shown in Fig.~\ref{fig:fig7}{\color{red}(b)}] in the MC scenario is around 8 times, and 2 orders of magnitude larger than ones in the GL and HL scenarios, respectively. Finally, the cross section of the process $e^{+}e^{-}\to\gamma\gamma \to \widetilde{\chi}_{2}^{0}\widetilde{\chi}_{2}^{0}$ [shown in Fig.~\ref{fig:fig7}{\color{red}(c)}] in the GL scenario is roughly 7 and 15 times larger than ones in the MC and HL scenarios, respectively.

It is well known that the neutralino/chargino masses, the relevant mixing matrices, and the couplings of the neutralino/chargino to other particles depend on the gaugino mass parameter $M_{2}$ and Higgsino mass parameter $\mu$.
\begin{figure}[hbpt]
    \begin{center}
\includegraphics[scale=0.47]{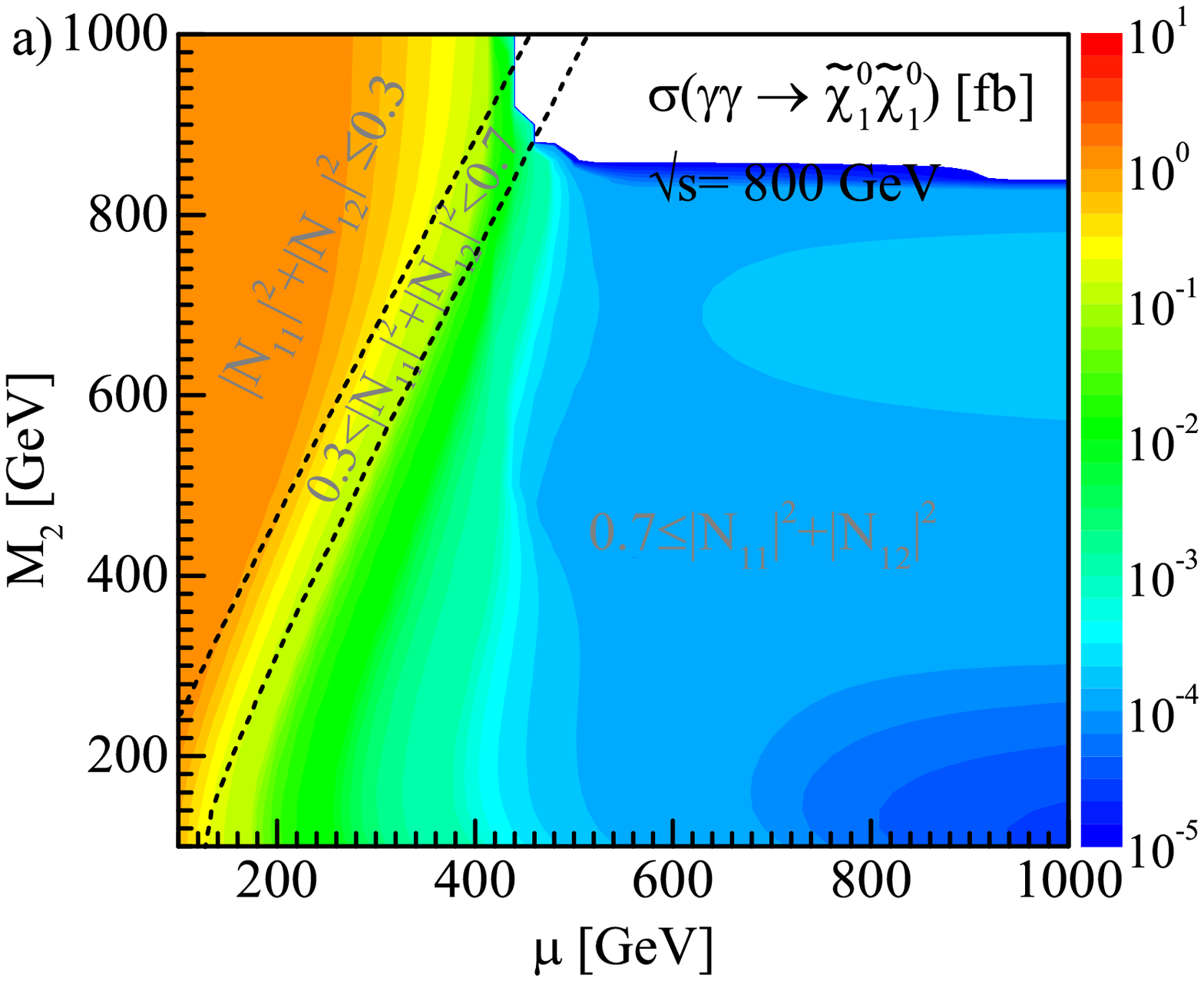}
\includegraphics[scale=0.47]{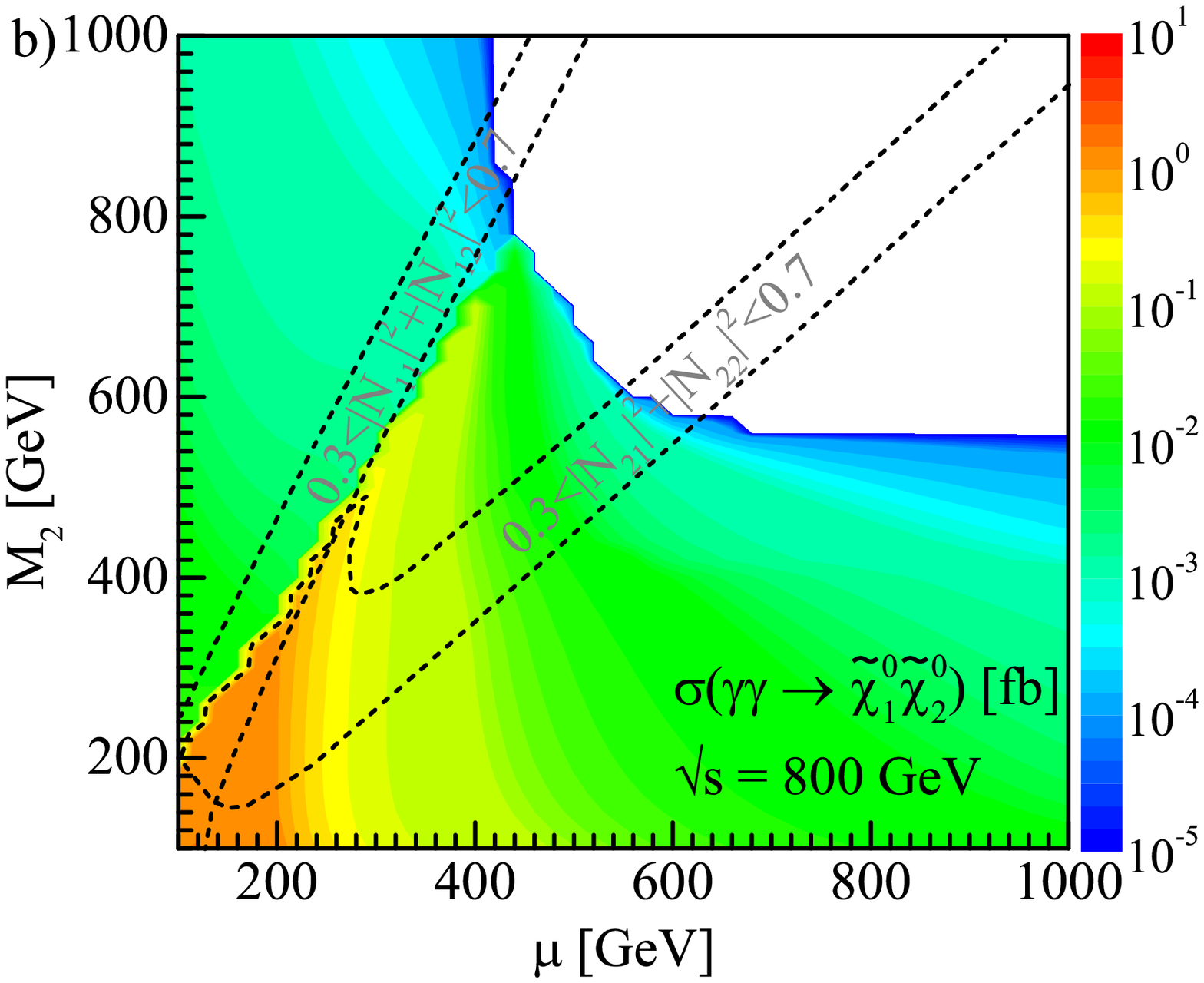}
\includegraphics[scale=0.47]{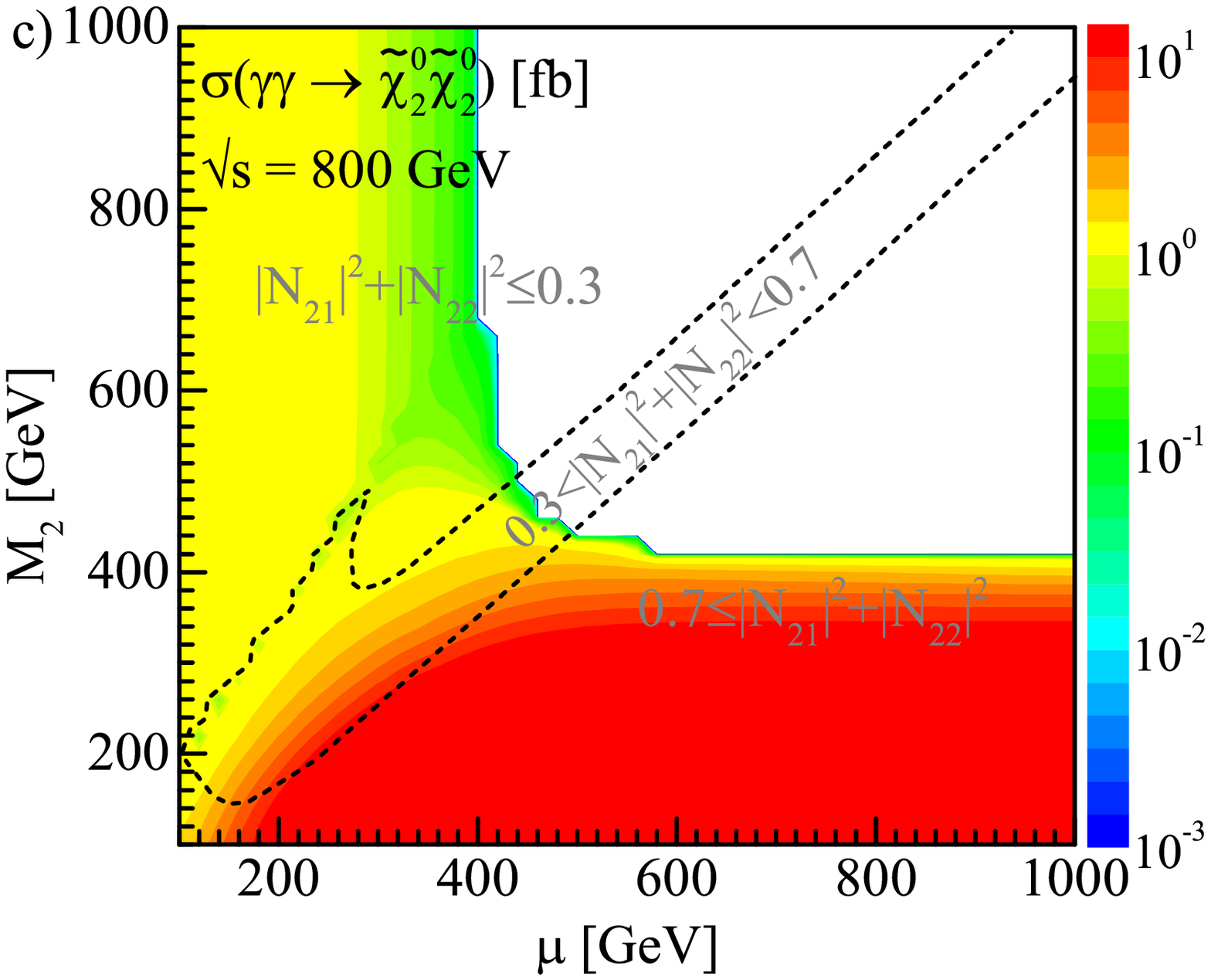}
      \end{center}
\caption{(color online). Contour plots of the total cross section of the process  $\gamma\gamma\to\widetilde{\chi}_{i}^0\widetilde{\chi}_{j}^0$ for (a) $i=j=1$, (b) $i=1,j=2$ and (c) $i=j=2$ as a 2D function of  the gaugino mass parameter $M_2$ and the Higgsino mass parameter $\mu$  at $\sqrt{\hat{s}}=800$ GeV for fixed values of $\tanb=10$ and $M_1=\frac{5}{3}M_2 \tan^2\theta_W$. The colored bins indicate the basic size of the total cross sections (in fb) in the scan region, and  the white areas represent parameter regions where production of a given pair of neutralinos is not kinematically accessible.}
\label{fig:fig8}
\end{figure}
Hence, the dependence of the cross section on these parameters can yield interesting information about the neutralinos/charginos. More importantly, by investigating the behavior with the these parameters, one can search for regions of the parameter space where the enhancement of cross section is large enough to be detectable at the ILC. In this context, we investigate the effect of these parameters on the integrated cross section of the neutralino pair production via photon-photon collisions ($\gamma\gamma\to\widetilde{\chi}_{i}^0\widetilde{\chi}_{j}^0$) in the $M_{2}$-$\mu$ mass plane by varying these parameters in the range from 100 to 1000 GeV in steps of 20 GeV at a center-of-mass energy of $\gamma\gamma$ $\sqrt{\hat{s}}=800$ GeV for $\tanb=$ 10, as depicted in Figs.~\ref{fig:fig8}{\color{red}(a)}-\ref{fig:fig8}{\color{red}(c)}.  Additionally, the $M_{2}$-$\mu$ mass plane is divided into regions based on the numerical values of the quantity $|N_{i1}|^2+|N_{i2}|^2$ such as the inequality $|N_{i1}|^2+|N_{i2}|^2\leq0.3$ representing the Higgsino-like region, the quantity $0.7\leq|N_{i1}|^2+|N_{i2}|^2$ representing the gauginolike region, and  the quantity $0.3<|N_{i1}|^2+|N_{i2}|^2<0.7$ corresponding to the mixture-case region for $i=1,2$. In these figures, the white regions mean that the center-of-mass energy is not large enough to produce a given final state of the neutralino pair $\widetilde{\chi}_{i}^0\widetilde{\chi}_{j}^0$. We see that the cross section of the $\gamma\gamma\to\neutralino_{i}\neutralino_{j}$ mostly increases with
increasing of $M_{2}$ and decreasing $\mu$ for the case of $i=1,j=1$, with a decrease of both $M_{2}$ and $\mu$ for the case of $i=1,j=2$ and with decreasing $M_{2}$ and increasing $\mu$ for the case of $i=2,j=2$. Particularly, the cross section reaches its maximum values in the region bounded by $|N_{11}|^2+|N_{12}|^2\leq0.3$ for $\gamma\gamma\to\neutralino_{1}\neutralino_{1}$ and $0.7\leq|N_{21}|^2+|N_{22}|^2$ for $\gamma\gamma\to\neutralino_{2}\neutralino_{2}$ into the scan region. On the other hand, the cross section of the process $\gamma\gamma\to\neutralino_{1}\neutralino_{2}$ usually reaches its maximum values in the intersection of the regions bounded by $0.3<|N_{11}|^2+|N_{12}|^2<0.7$ and $0.3<|N_{21}|^2+|N_{22}|^2<0.7$.
One can conclude that pure Higgsino couplings dominate in the case of $i=1,j=1$, whereas pure gaugino couplings dominate in the case of $i=2,j=2$ for $\gamma\gamma\to\neutralino_{i}\neutralino_{j}$. Also, a nearly equal mixture of gaugino and Higgsino couplings dominate in process $\gamma\gamma\to\neutralino_{1}\neutralino_{2}$. These results are in agrement with dominant scenarios which appear in the center-of-mass energy and $\tanb$ dependencies of the cross section. As a consequence,
it is clear that the cross section strongly depends on the parameters $M_{2}$ and $\mu$.

Two alternative projects for linear $e^{+}e^{-}$ collider are presently under consideration: the ILC and CLIC. The ILC is based on superconducting technology in the TeV range, while the CLIC is based on the novel approach of two-beam acceleration to extend linear colliders into the multi-TeV range. The choice will be based on the respective maturity of each technology and on the physics requests derived from the LHC physics results when available. The CLIC aims at multi-TeV collision energy with high luminosity about $\mathcal{L}_{e^+ e^-}\sim8\cdot10^{34} $cm$^{-2}$s$^{-1}$ \cite{Ellis,Asner,CLIC1}. Furthermore, a photon-photon collider is one option of the ILC in the center-of-mass energy $\sqrt{s}=250-1000$ GeV with an integrated luminosity of $100$ fb$^{-1}$ per year. This collider is expected to be upgradeable to 1 TeV with total integrated luminosity up to $300$ fb$^{-1}$ yearly. At the ILC, once the kinematical threshold is crossed for the neutralino/chargino pair production, the signals of neutralinos and charginos may be separated with O(1)$\%$ measurements for the couplings and the mass resolution \cite{Berggren,Pandita}. Our results indicate that the cross section reaches its largest values when the lightest neutralino has a dominant Higgsino component and/or the next-to-lightest neutralino has a dominant gaugino component. Moreover, the sizes of the total cross sections are at a visible level of $10^{-1}$ to $10^{1}$ fb. A comparison of our results with the technical data of the ILC, hence, leads us to conclude that our proposed channels are likely observable at the ILC.

Note that the decay products of a Higgsino-like next-to-lightest neutralino are different from those of a gauginolike one. In our scenarios, the lightest sleptons are even heavier than the neutralinos, thus leading to branching fractions of 100$\%$ for decay $\neutralino_2 \to Z \neutralino_1$ in the GL scenario and 59$\%$ for decay $\neutralino_2 \to h \neutralino_1$ and 41$\%$ for $\neutralino_2 \to Z \neutralino_1$ in the MC scenario\footnote{We have calculated the branching ratios with the help of the program the \texttt{CompHEP}  (version 4.5.2) \cite{CompHEP}.}. In the HL scenario, two-body decays for the neutralino $\neutralino_2$ are not kinematically allowed since the mass difference between the LSPs and $\neutralino_2$ are typically small. However, if the decays of neutralino $\neutralino_{2}$ with a significant Higgsino content into third-family quark-squark pairs were opened, they would be greatly enhanced by the top-quark Yukawa coupling. Consequently, the detection of Higgsino-like or gauginolike next-to-lightest neutralinos requires different cuts implying different efficiencies and acceptances.

\section{Conclusion}\label{sec:conclusion}
In this work, we have evaluated the total cross section of the neutralino pair production via photon-photon fusion, which arises for the first time at one-loop level, in electron-positron collisions at a linear collider. We have investigated numerically the effects of the the center-of-mass energy, the tan$\beta$, and the $M_2$-$\mu$ mass plane on the total cross section for three different scenarios, the HL, GL, and MC, in the framework of the MSSM. Furthermore, we have analyzed the contribution of triangle-type, bubble-type, box-type, and quartic-coupling diagrams to the total cross section.

Our numerical results affirm that the total cross sections remarkably increase due to resonance effects of diagrams involving the intermediate neutral Higgs boson $H^0$ or $A^0$. When considered from this point of view, investigation of the neutralino pair production via photon-photon fusion is significant for the experimental and phenomenological works in connection with the neutral Higgs bosons at the linear colliders. The basic size of the total cross sections is seen to differ by several orders of magnitude depending on the gaugino/Higgsino fractions of the produced neutralinos. The total cross section is mostly dominated by the Higgsino-like scenario for the process $e^{+}e^{-}\to\gamma\gamma\to\widetilde{\chi}_{1}^0\widetilde{\chi}_{1}^0$, the micture-case scenario for the process $e^{+}e^{-}\to\gamma\gamma\to\widetilde{\chi}_{1}^0\widetilde{\chi}_{2}^0$, and gauginolike scenario for the process $e^{+}e^{-}\to\gamma\gamma\to\widetilde{\chi}_{2}^0\widetilde{\chi}_{2}^0$. In particular, the total cross section of the production of $\widetilde{\chi}_{2}^0\widetilde{\chi}_{2}^0$ in the gauginolike scenario reaches a value of 2.23 fb at $\sqrt{s}= 900$ GeV and may be more observable than others at the ILC. Analysis of the individual contributions of different types of diagrams to the total cross section shows that the box-type diagrams make the main contribution to the total cross section for all of the considered scenarios due to a destructive interference occurring between the triangle- and bubble-type diagrams. It should be noted that further enhancement of the cross section can be seen locally as a result of the threshold effects occurring because of relatively light intermediate states in box-type diagrams.

Furthermore, it is quite obvious that the total cross section is not very sensitive with respect to variation of the tan$\beta$ (particularly for intermediate to larger values of tan$\beta$), but it is strongly dependent on the parameters $M_{2}$ and $\mu$ as expected. From the parameters $M_2$ and $\mu$ dependence of the cross section, we can conclude that pure Higgsino couplings dominate in the case of $i=1,j=1$, whereas pure gaugino couplings dominate in the case of $i=2,j=2$ for $\gamma\gamma\to\neutralino_{i}\neutralino_{j}$. Consequently, we can point out that the enhancement of the cross section appears when the lightest neutralino $\widetilde{\chi}_{1}^0$ has a dominant Higgsino component and/or the next-to-lightest neutralino $\widetilde{\chi}_{2}^0$ has a dominant gaugino component.

It should be underlined that a model-independent and precise measurement of the cross section for neutralino pair production via photon-photon fusion would be possible at a linear $e^{+} e^{-}$ collider such as the ILC or the CLIC, and hence the results obtained in this work will be helpful in detecting SUSY signals, in determining the basic SUSY parameters, and in deriving more precise bounds on neutralino/chargino masses and the corresponding couplings.

\begin{acknowledgments}
M. D. would like to thank T. Hahn for sending the updated version of CompAZ.F, which is used for computing photon luminosity. The Feynman diagrams in this work have been drawn with the help of the program \texttt{JaxoDraw 2.0} \cite{JaxoDraw, JaxoDraw2}.
\end{acknowledgments}


\begin{thebibliography}{99}
\bibitem{Nilles} H. P. Nilles, 
  \href{http://dx.doi.org/10.1016/0370-1573(84)90008-5}{Phys. Rep.~{\bf 110}, 1 (1984)}.
\bibitem{Haber} H. E. Haber and G. L. Kane, 
    \href{http://dx.doi.org/10.1016/0370-1573(85)90051-1}{Phys. Rep.~{\bf 117}, 75 (1985)}.
\bibitem{wss} H. Baer and X. Tata, {\it Weak Scale Supersymmetry: From Superfields to Scattering Events}, (Cambridge University Press, Cambridge, England, 2006).
\bibitem{DM} J. L. Feng, 
    \href{http://dx.doi.org/10.1146/annurev-astro-082708-101659}{Ann. Rev. Astron. Astrophys.~\textbf{48}, 495 (2010)}. 
\bibitem{Higgs_ATLAS} G.~Aad \textit{et al.}  (ATLAS Collaboration), 
    \href{http://dx.doi.org/10.1016/j.physletb.2012.08.020}{Phys. Lett. B} \href{http://dx.doi.org/10.1016/j.physletb.2012.08.020}{{\bf 716}, 1 (2012)}. 
\bibitem{Higgs_CMS} S.~Chatrchyan \textit{et al.} (CMS Collaboration), 
    \href{http://dx.doi.org/10.1016/j.physletb.2012.08.021}{Phys. Lett. B} \href{http://dx.doi.org/10.1016/j.physletb.2012.08.021}{{\bf 716}, 30 (2012)}. 
\bibitem{Heinemeyer} S. Heinemeyer, O. Stal, and G. Weiglein, 
    \href{http://dx.doi.org/10.1016/j.physletb.2012.02.084}{Phys. Lett.B} \href{http://dx.doi.org/10.1016/j.physletb.2012.02.084}{{\bf 710}, 201 (2012)}.
\bibitem{Scopel} S. Scopel, N. Fornengo, and A. Bottino,  
    \href{http://link.aps.org/doi/10.1103/PhysRevD.88.023506}{Phys. Rev. D} \href{http://link.aps.org/doi/10.1103/PhysRevD.88.023506}{{\bf 88}, 023506 (2013)}.
\bibitem{Djouadi} A. Djouadi,  
    \href{http://dx.doi.org/10.1140/epjc/s10052-013-2704-3}{Eur. Phys. J. C~{\bf 74}, 2704 (2014)}.

\bibitem{Fayet} P.~Fayet, 
    \href{http://dx.doi.org/10.1016/0370-2693(77)90852-8}{Phys. Lett. B~{\bf 69}, 489 (1977)}; G. R.~Farrar and P.~Fayet, 
     \href{http://dx.doi.org/10.1016/0370-2693(78)90858-4}{Phys. Lett. B~{\bf 76}, 575  (1978)}.
\bibitem{Jungman} G.~Jungman, M.~Kamionkowski, and K.~Griest,
     \href{http://dx.doi.org/10.1016/0370-1573(95)00058-5}{Phys. Rep. {\bf 267}, 195 (1996)}.
\bibitem{Griest} K. Griest and M. Kamionkowski, 
    \href{http://dx.doi.org/10.1016/S0370-1573(00)00021-1 }{Phys. Rep.~{\bf 333},}  \href{http://dx.doi.org/10.1016/S0370-1573(00)00021-1 }{167 (2000)}.
\bibitem{Arnowitt} R. Arnowitt and P. Nath, 
    \href{http://link.aps.org/doi/10.1103/PhysRevD.54.2374}{Phys. Rev. D~{\bf 54}, 2374 (1996)}.
\bibitem{sqgl_ATLAS1} G.~Aad \textit{et al.}  (ATLAS Collaboration), 
   \href{http://dx.doi.org/10.1007/JHEP10(2015)054}{J. High Energy} \href{http://dx.doi.org/10.1007/JHEP10(2015)054}{Phys.~{\bf 10} (2015) 054 }.
\bibitem{sqgl_ATLAS2} M.~Aaboud \textit{et al.}  (ATLAS Collaboration), 
   \href{http://dx.doi.org/10.1140/epjc/s10052-016-4184-8}{Eur. Phys.} \href{http://dx.doi.org/10.1140/epjc/s10052-016-4184-8}{J. C {\bf 76}, 392 (2016)}.
\bibitem{sqgl_CMS1} S.~Chatrchyan \textit{et al.} (CMS Collaboration), 
    \href{http://link.aps.org/doi/10.1103/PhysRevD.90.112001}{Phys. Rev. D} \href{http://link.aps.org/doi/10.1103/PhysRevD.90.112001}{{\bf 90}, 112001 (2014)}.
\bibitem{sqgl_CMS2} V.~Khachatryan \textit{et al.} (CMS Collaboration), 
    Report No. \href{https://cds.cern.ch/record/2160221}{CMS-SUS-15-010 (2016)}, \href{https://arxiv.org/abs/1606.03577}{arXiv:1606.03577 [hep-ex]}.
\bibitem{Chan} K. L. Chan, U. Chattopadhyay, and P. Nath, 
    \href{http://dx.doi.org/10.1103/PhysRevD.58.096004}{Phys. Rev. D~{\bf 58}, 096004 (1998)}. 
\bibitem{Abdallah} J. Abdallah \textit{et al.} (DELPHI Collaboration), 
    \href{http://dx.doi.org/10.1140/epjc/s2003-01355-5}{Eur. Phys.} \href{http://dx.doi.org/10.1140/epjc/s2003-01355-5}{J. C~{\bf 31}, 421 (2003)}.
\bibitem{PDG} K. A.~Olive \textit{et al.} (Particle Data Group), 
    \href{http://stacks.iop.org/1674-1137/38/i=9/a=090001}{Chin. Phys. C} \href{http://stacks.iop.org/1674-1137/38/i=9/a=090001}{{\bf 38}, 090001 (2014)}.
\bibitem{Han} T.~Han, S.~Padhi, and S.~Su, 
   \href{http://link.aps.org/doi/10.1103/PhysRevD.88.115010}{Phys. Rev. D~{\bf 88},} \href{http://link.aps.org/doi/10.1103/PhysRevD.88.115010}{115010 (2013)}.
\bibitem{HBaer} H.~Baer, V.~Barger, P.~Huang, D.~Mickelson, A.~Mustafayev, W.~Sreethawong, and X.~Tata, 
   \href{http://link.aps.org/doi/10.1103/PhysRevLett.110.151801}{Phys. Rev. Lett.~{\bf 110}, 151801 (2013)}.
\bibitem{Demirci3} M.~Demirci and A. I.~Ahmadov, 
    \href{http://link.aps.org/doi/10.1103/PhysRevD.89.075015}{Phys. Rev. D~{\bf 89},}  \href{http://link.aps.org/doi/10.1103/PhysRevD.89.075015}{075015 (2014)}, \href{http://arxiv.org/abs/1404.0550}{arXiv:1404.0550 [hep-ph]}.

\bibitem{ILC1} T. Behnke \textit{et al.}, {\it The International Linear Collider Technical Design Report - Volume 1: Executive Summary}, 	Report No. ILC-REPORT-2013-040,
   \href{http://arxiv.org/abs/1306.6327}{arXiv:1306.6327 [physics.acc-ph]}.
\bibitem{ILC2} H.~Baer \textit{et al.}, {\it Physics Chapter of the ILC Detailed Baseline Design Report}, Report No. ILC-REPORT-2013-040,
   \href{http://arxiv.org/abs/1306.6352}{arXiv:1306.6352 [hep-ph]}.

\bibitem{eeLC} G. Moortgat-Pick \textit{et al.},  
    \href{http://dx.doi.org/10.1140/epjc/s10052-015-3511-9}{Eur. Phys. J. C~{\bf 75}, 371 (2015)}.
\bibitem{CLILC1} L. Linssen, A. Miyamoto, M. Stanitzki, and H. Weerts, {\it Physics and Detectors at CLIC: CLIC Conceptual Design Report}, Report No. CERN-2012-003,
   \href{http://arxiv.org/abs/1202.5940}{arXiv:1202.5940 [physics.ins-det]}.
\bibitem{HBaer2} H.~Baer, M.~Berggren, J. List, M. M.~Nojiri, M.~Perelstein, A.~Pierce, W. Porod and T. Tanabe, 
    \href{http://arxiv.org/abs/1307.5248}{arXiv:1307.5248 [hep-ph]}.

\bibitem{Fei} Zhou Fei, Ma Wen-Gan, Jiang Yi, and Han Liang, 
    \href{http://link.aps.org/doi/10.1103/PhysRevD.62.115006}{Phys. Rev. D~{\bf 62}, 115006 (2000)}.
\bibitem{Gounaris} G. J. Gounaris, J. Layssac, P. I. Porfyriadis, and F. M. Renard,  
    \href{http://dx.doi.org/10.1140/epjc/s2003-01520-x}{Eur. Phys. J. C~{\bf 32}, 561 (2004)}.
\bibitem{Nasuf} N. Sonmez, 
    \href{http://link.aps.org/doi/10.1103/PhysRevD.91.085021}{Phys. Rev. D~{\bf 91}, 085021 (2015)}.

\bibitem{Feynarts} J.~K\"ublbeck, M. B\"ohm, and A. Denner, 
    \href{http://dx.doi.org/10.1016/0010-4655(90)90001-H}{Comput. Phys. Commun.~{\bf 60}, 165 (1990)};
     T.~Hahn, 
     \href{http://dx.doi.org/10.1016/S0010-4655(01)00290-9}{Comput. Phys. Commun.~{\bf 140}, 418 (2001)}. 
\bibitem{Telnov} V. I. Telnov, 
    \href{http://dx.doi.org/10.1016/0168-9002(90)91826-W}{Nucl. Instrum. Methods Phys. Res., Sect. A} \href{http://dx.doi.org/10.1016/0168-9002(90)91826-W}{~{\bf 294}, 72 (1990)}.
\bibitem{Hahn} T.~Hahn and C. Schappacher, 
     \href{http://dx.doi.org/10.1016/S0010-4655(01)00436-2}{Comput. Phys.} \href{http://dx.doi.org/10.1016/S0010-4655(01)00436-2}{Commun.~{\bf 143}, 54 (2002)}. 
\bibitem{loop} T.~Hahn and M. Perez-Victoria, 
    \href{http://dx.doi.org/10.1016/S0010-4655(98)00173-8}{Comput. Phys.} \href{http://dx.doi.org/10.1016/S0010-4655(98)00173-8}{Commun.~{\bf 118}, 153 (1999)}. 
\bibitem{FeynHiggs} S.~Heinemeyer, W. Hollik, and G. Weiglein, 
    \href{http://dx.doi.org/10.1016/S0010-4655(99)00364-1}{Comput.} \href{http://dx.doi.org/10.1016/S0010-4655(99)00364-1}{Phys. Commun.~{\bf 124}, 76 (2000)}. 

\bibitem{Compaz} A. F. Zarnecki, 
    \href{http://arxiv.org/abs/hep-ex/0207021}{Acta Physica Pol. B~{\bf 34}, 2741 (2003)}, \href{http://arxiv.org/abs/hep-ex/0207021}{arxiv:hep-ex/0207021}.

\bibitem{Demirci1} A. I.~Ahmadov and M.~Demirci, 
    \href{http://dx.doi.org/10.1142/S0217751X13500772}{Int. J. Mod. Phys. A}~{\color{blue}{\bf28},} \href{http://dx.doi.org/10.1142/S0217751X13500772}{ 1350077 (2013)},  \href{http://arxiv.org/abs/1307.3779v1}{arXiv:1307.3779 [hep-ph]}.
\bibitem{Demirci2} A. I.~Ahmadov and M.~Demirci, 
    \href{http://dx.doi.org/10.1103/PhysRevD.88.015017}{Phys. Rev. D~{\bf 88},} {\color{blue}015017 (2013)}, \href{http://arxiv.org/abs/1307.3777}{arXiv:1307.3777 [hep-ph]}.
\bibitem{Pandita} P. N. Pandita and M. Patra, 
    \href{http://link.aps.org/doi/10.1103/PhysRevD.88.055018}{Phys. Rev. D} \href{http://link.aps.org/doi/10.1103/PhysRevD.88.055018}{~{\bf 88}, 055018 (2013)}.
\bibitem{Ellis} J. R. Ellis and I. Wilson, 
    \href{http://dx.doi.org/10.1038/35053224}{Nature} {\color{blue}(London)}~\href{http://dx.doi.org/10.1038/35053224}{{\bf 409}, 431 (2001)}.
\bibitem{Asner} D. Asner, H. Burkhardt, A. De Roeck, J. Ellis, J. Gronberg, S. Heinemeyer, M. Schmitt, D. Schulte, M. Velasco, and F. Zimmermann,  
    \href{http://dx.doi.org/10.1140/epjc/s2002-01113-3}{Eur. Phys. J. C~{\bf 28}, 27 (2003)}.
\bibitem{CLIC1} E. Accomando \textit{et al.}~(CLIC Physics Working Group), {\it Physics at the CLIC Multi-TeV Linear Collider}, Report No. CERN-2004-005,
   \href{https://arxiv.org/abs/hep-ph/0412251}{arXiv:hep-ph/0412251}.
\bibitem{Berggren} M. Berggren, F. Br{\"u}mmer, J. List, G. Moortgat-Pick, T. Robens, K. Rolbiecki and H. Sert,  
    \href{http://dx.doi.org/10.1140/epjc/s10052-013-2660-y}{Eur. Phys. J. C~{\bf 73}, 2660 (2013)}.

\bibitem{CompHEP} E. Boos, V. Bunichev, M. Dubinin, L. Dudko, V. Edneral, V. Ilyin, A. Kryukov, V. Savrin, A. Semenov, and A. Sherstnev,  
    {\color{blue}Nucl.} \href{http://dx.doi.org/10.1016/j.nima.2004.07.096}{Instrum. Methods Phys. Res., Sect. A~{\bf 534}, 250 (2004)}.
\bibitem{JaxoDraw} D. Binosi and L. Theussl, 
    \href{http://dx.doi.org/10.1016/j.cpc.2004.05.001}{Comput. Phys.} \href{http://dx.doi.org/10.1016/j.cpc.2004.05.001}{Commun.~{\bf 161}, 76 (2004)}.
\bibitem{JaxoDraw2} D. Binosi, J. Collins, C. Kaufhold, and L. Theussl, 
    \href{http://dx.doi.org/10.1016/j.cpc.2009.02.020}{Comput. Phys. Commun.~{\bf 180}, 1709 (2009)}. 

\end{thebibliography}
\end{document}